\documentclass[conference,compsoc]{IEEEtran}

\ifCLASSOPTIONcompsoc
  \usepackage[nocompress]{cite}
\else
  \usepackage{cite}
\fi

\usepackage{footnote}
\usepackage{url}
\usepackage{hyperref}
\usepackage{adjustbox}
\usepackage{hhline}
\usepackage{graphicx}
\usepackage{array}
\usepackage[font=scriptsize,labelfont=sf,textfont=sf,lofdepth,lotdepth]{subfig}
\usepackage{multirow}
\usepackage{cleveref}
\usepackage{epstopdf} 
\usepackage[ruled,vlined,linesnumbered,resetcount]{algorithm2e}
\usepackage{algpseudocode}

\usepackage{mathtools}
\usepackage{amssymb}
\usepackage{amsmath}
\usepackage{siunitx}

\usepackage{tikz}
\usepackage{forest}
\usetikzlibrary{arrows.meta}

\def\nicefrac#1#2{\leavevmode%
    \raise.5ex\hbox{\small #1}%
    \kern-.1em/\kern-.15em%
    \lower.25ex\hbox{\small #2}}

\usepackage{booktabs}
\newcommand\myurl[2]{\url{#1}}

\IEEEoverridecommandlockouts

\begin{document}
\title{Trollslayer: Crowdsourcing and Characterization of Abusive Birds in Twitter}


\author{\IEEEauthorblockN{\'{A}lvaro Garc\'{i}a-Recuero *\thanks{* The first and main author of this work developed Trollslayer at INRIA since 2015, published as \href{https://doi.org/10.1109/SNAMS.2018.8554898}{https://doi.org/10.1109/SNAMS.2018.8554898}.
}}
\IEEEauthorblockA{Queen Mary University of London\\
alvaro.garcia-recuero@qmul.ac.uk}
\and
\IEEEauthorblockN{Aneta Morawin}
\IEEEauthorblockA{INRIA Rennes, France\\
firstname.secondname@inria.fr}
\and
\IEEEauthorblockN{Gareth Tyson}
\IEEEauthorblockA{Queen Mary University of London\\
g.tyson@qmul.ac.uk}
}

\maketitle

\begin{abstract}
As of today abuse is a pressing issue to participants and administrators of Online Social Networks (OSN). Abuse in Twitter can spawn from arguments generated for influencing outcomes of a political election, the use of bots to automatically spread misinformation, and generally speaking, activities that {\em deny}, {\em disrupt}, {\em degrade} or {\em deceive} other participants and, or the network. Given the difficulty in finding and accessing a large enough sample of abuse ground truth from the Twitter platform, we built and deployed a custom crawler that we use to judiciously collect a new dataset from the Twitter platform with the aim of characterizing the nature of abusive users, a.k.a abusive ``birds'', in the wild. We provide a comprehensive set of features based on users' attributes, as well as social-graph metadata. The former includes metadata about the account itself, while the latter is computed from the social graph among the sender and the receiver of each message. Attribute-based features are useful to characterize user's accounts in OSN, while graph-based features can reveal the dynamics of information dissemination across the network. In particular, we derive the Jaccard index as a key feature to reveal the benign or malicious nature of directed messages in Twitter. To the best of our knowledge, we are the first to propose such a similarity metric to characterize abuse in Twitter.
\end{abstract}

\IEEEpeerreviewmaketitle
\section{Introduction} 
Users of OSN are exposed to abuse by other participants, who typically send their victims harmful messages designed to {\em deny}, {\em disrupt}, {\em degrade} and {\em deceive} among a few, as reported by top secret methods for online cyberwarfare in JTRIG~\cite{JTRIGs}. In Twitter, these practices have a non-negligible impact in the manipulation of political elections~\cite{Ferrara2015}, fluctuation of stock markets~\cite{Bollen2011} or even promoting terrorism~\cite{twitter-suspension}. As of today, and in the current turmoil of fake news and hate speech, we require a global definition for ``abuse''. We find the above definition from JTRIG to be able to cover all types of abuse we find in OSN as of today. Secondly, to identify abuse the Twitter platform often relies on participants reporting such incidents of abuse. In other OSN as Facebook this is also the case, as suggested by the large number of false positives encountered by~\cite{boshmafbots2011} in the Facebook Immune System~\cite{immune}. In addition, Twitter suspending abusive participants can be seen as censorship, as it effectively limits free speech of users in the Internet. Finally, user's privacy is today an increasing concern for users of large OSN. Privacy often clashes with efforts for reducing abuse in these platforms~\cite{FrenchCourt} because even disclosing metadata that holds individuals accountable in such cases violates the fundamental right to privacy according to the Universal Declaration of Human Rights~\cite{UN}. In the same vein, and back to the Twitter platform, we observe a constant trading of individuals' privacy for granting governments access to private metadata. This endangers citizens well-being and puts them into the spotlight for law enforcement to charge them with criminal offenses, even when no serious criminal offense has been committed~\cite{caution:2012}.

The main contribution of this paper is a large-scale study of the dynamics of abuse in a popular online social micro-blogging media platform, Twitter. For that, we collect a dataset where we annotate a subset of the messages received by potential victims of abuse in order to characterize and assess the prevalence of such malicious messages and participants. Also, we find it revealing to understand how humans agree or not in what represents abuse during the crowd sourcing. In summary, the aim of the study is to answer the following research questions (RQ):

\textbf{RQ.1:}
Can we obtain relevant abuse ground truth from a large OSN such as Twitter using BFS (Bread-First-Search) sampling for data collection and crowd-sourcing for data annotation? We show statistics about the dataset collected and the annotated dataset respectively.

\textbf{RQ.2:}
Does it make sense to characterize abuse from a victim's point of view? We provide a list of user attributes (local) and graph-based (global) features that can characterize abusive behavior.

\textbf{RQ.3:}
What are the dynamics of abusive behavior? Does it appear as an isolated incident or is it somehow organized? We show that the source of several messages comes from an automated social media scheduling platform that redirects Twitter users to a doubtful site about a fund-raising campaign for a charity (considered as deceive in the abuse definition we employ).

\section{Victim-Centric Methodology}
In order to collect data from Twitter we adapt the usual BFS for crawling social media and start crawling data from a sufficiently representative number of accounts for our measurement, which we we call the victims' seed set. The first half of accounts are likely victims, chosen independently of any sign or trace of abuse in their public Twitter timeline 
in order to account for randomness in the measurements. The second half is selected based in their public timeline containing traces or likelihood of abuse, namely potential victims of abuse. Therefore, we define the seed set as made up of potential victims and likely victims. We then bootstrap our crawler, following the recursive procedure in Algorithm~\ref{algo:bfs}, which collects messages directed towards each of the seeds. If a message is directed towards or mentioning two or more victims, we consider it several times for the same message sender but with different destinations. We also collect the subscription and subscriber accounts of sender and receiver in the Twitter social graph, namely follower and followee relationships.

\subsection{Data model}\label{datamodel}

Consider a seed set of nodes for forming a graph $\mathcal{G}_s$=$(\mathcal{V}_s, \mathcal{E}_s)$ containing the nodes in the seed set (victims) and their potential perpetrators as the two entities defining the edge relationships in $\mathcal{E}_s$. Given that $\mathcal{G}_s$ is a directed graph made of vertices $(\mathcal{V}_s)$ and edges $(\mathcal{E}_s)$ making up a connection or defining a message sent among a pair of nodes $(u,v)$, we derive two specialized directed graphs with their corresponding relationships, messaging or social follow in the network.

Firstly, let $\mathcal{G}_f$=$(\mathcal{V}_f, \mathcal{E}_f)$ be a directed graph of social relationships where the vertices $\mathcal{V}_f$ represent users and a set of directed edges $\mathcal{E}_f$ representing subscriptions:

\begin{equation*}
  \mathcal{E}_f \coloneqq \{ (u, v) \mid u \textrm{ publicly follows } v\}
\end{equation*}

Secondly, let $\mathcal{G}_m$=$(\mathcal{V}_m, \mathcal{E}_m)$ be a directed messaging multi-graph with a set of users as vertices $\mathcal{V}_m$, and a set of directed edges representing messages sent by user $u$ mentioning user $v$:

\begin{equation*}
  \mathcal{E}_m \coloneqq \{ (u, v) \mid u \textrm{ messages } v\ \textrm{with a public mention} \} 
\end{equation*}

$\mathcal{E}_m$ models the tweets that are shown to users with or without explicit subscription by the recipient to the sender. Thus, these messages represent a vector for abusive behavior.

To bootstrap our crawler, we start with the mentioned \emph{seed set} and run an adapted and recursive \emph{bounded breath-first-search} (bBFS) procedure on the Twitter input seeds to cover up to a maximum depth {\em maxdepth} we pass as parameter to it. In Algorithm~\ref{algo:bfs} we summarize the operational mode of {\em bBFS}.


\alglanguage{pseudocode}
\begin{algorithm}
 \SetKwInOut{Input}{input}
 \SetKwInOut{Parameter}{param}
 \SetKwInOut{Output}{output}
 \SetKwInOut{Function}{function}
 \SetKwInOut{Global}{global}
 \SetKwData{Depth}{$depth$}
 \SetKwData{Add}{$add$}
 \SetKwData{CurrentId}{$currentid$}
 \SetKwData{MaxFollows}{$maxfollows$}
 \SetKwData{Parent}{$parent$}
 \SetKwData{Distance}{$distance$}
 \SetKwData{Head}{$head$}
 \SetKwFunction{Pop}{pop}
 \SetKwFunction{Push}{push}
 \SetKwFunction{Append}{append}
 \SetKwFunction{Reassign}{reassign}
 \SetKwFunction{GetFollowersU}{$\mathcal{F}(u)$}
 \SetKwFunction{GetFollowersV}{$\mathcal{F}(v)$}
 \SetKwFunction{Map}{map()}
 \SetKwFunction{Set}{set}
 \SetKwFunction{Append}{append}
 \SetKwFunction{Insert}{insert}
 \SetKwFunction{proc}{bBFS}%
 \SetKwProg{myproc}{Procedure}{}{end}
 \SetKwComment{tcp}{//}{}

\BlankLine
\Input{$seeds$ is the set of potential victims}
\Input{$newseeds$ is the set of new seeds from follower of potential victims}

\BlankLine
\Parameter{$maxFollows$ controls the number of followers}
\Parameter{$maxdepth$ parameter controls the crawling depth}

\BlankLine
\Function{\(\mathcal{F}(x) \coloneqq
   \{y \mid (y, x) \in \mathcal{E}\)\} gets followers of node \(x \in
   \mathcal{G}\)}
\Function{\(Add(x, y) \coloneqq
   {(x, y) \mid y \in \mathcal{F}(x)}\)~maps follower $y$ to parent
   item $x \in \mathcal{G}$}   

\BlankLine
\Output{$\mathcal{G}$=$(\mathcal{V}$, $\mathcal{E})$}

\BlankLine
\myproc{\proc{$seeds, depth, maxdepth$}}{
\If{not $seeds$}{\Return $\mathcal{G}$}
$newseeds$ \(\gets\) \{\} \\
\While{seeds}{
  \(u\) \(\gets\) seeds.\Pop \\
  \If{$|\GetFollowersU|$ \(\leq\) \MaxFollows \bf{and} \Depth \( \leq \) $maxdepth +1$}
  {
    \While{$\GetFollowersU$}
    {
    \ForEach{\(v\) \(\in\) $\GetFollowersU$}
      {
      \(\mathcal{B} \gets Add(u, v)\) \\
      $\mathcal{G}$ \(\gets\) $\mathcal{G}$ $\cup$ $\mathcal{B}$ \\
  	    \If{\Depth + 1 \( \leq \) $maxdepth$}
        {
        $newseeds$.\Append{\(v\)}
        }
      }
    } 
  }
} 
\Depth \(\gets\) \Depth + 1 \\
\Return \proc{$newseeds, $depth$, maxdepth$}
} 
\caption{Recursive bBFS procedure}\label{algo:bfs}
\end{algorithm}

\subsection{Boundaries of the data crawl}
The configuration of the crawler controls from where the crawl starts and puts some restrictions on where it should stop. The first one of such restrictions during the graph traversal is collecting incoming edges a.k.a followers in Twitter when the number does not exceed an upper bound, depending on the chosen {\em maxfollowers} as node popularity. Secondly, the followers must be within a maximum depth we call {\em maxdepth} in order to collect the related metadata in the graph belonging to them. 

For each node meeting the above constraints, we also collect user account metadata as well as their respective public timeline of messages metadata in Twitter; then we start crawling the followers of nodes at depth 1, and next depth 2 (followers of followers)and so on as set by the parameter mentioned. In our dataset, we never go any further than second degree followers to collect relationships among users in the social graph crawled. 

\subsection{Data annotation}\label{gt}
To annotate abuse we have developed an in-house crowd-sourcing platform, {\em Trollslayer} \footnote{\url{https://github.com/algarecu/trollslayer}}, where we enlisted ourselves and various colleagues to assist with the tedious effort of annotating abuse. However, we decide to enlarge our annotations with the support of a commercial crowd-sourcing platform named {\em Crowdflower}, where we spent around \$30 in credit using a student data for everyone pack. In the crowd sourcing process we account for scores collected from 156 crowd workers in {\em Crowdflower} and 7 trusted crowd workers in {\em Trollslayer}, accounting to 163 crowd workers overall. In these two platforms we display the same tweets and the same guidelines to crowd workers that annotate messages. Therefore, we are able to compute the global scores from both platforms on the same tweets to end up with at least 3 annotations per tweet inthe worst case.

\section{Dataset}
So far we have judiciously collected a dataset from Twitter to characterize abuse in Twitter. Using crowd workers we obtain abuse ground truth. Next we extract relatively simple features from the collected dataset. Given that the features are largely based on data that is available in the proximity of the potential victim, we aim to characterize the distribution of abuse in an online micro-blogging platform from the view of the victim. This also avoids the Big Data mining that can only be effectively performed by large micro-blogging service providers.
%

\subsection{Statistics}\label{sub:dataset-stats}
Table~\ref{table:crawl} shows statistics about the dataset collected such as the number of tweets directed toward the list of victims in our seed set. In total, we account for 1648 tweets directed to our seed set at depth 1. Then we show the same statistics organized by \emph{depth} in the recursive crawl performed to obtain the dataset. Note that for the purpose of the statistical analysis of the dataset and findings presented here, we will only take into consideration nodes for which the social graph has been fully collected. Due to Twitter Terms and Conditions (TTC) we plan to make available and public only the identifiers of the messages annotated but not the rest of the information associated to the message, graph or private information that identifies the crowd-workers.


\begin{table}[!ht]
  \begin{center}
  \begin{adjustbox}{max width=\columnwidth}
  \begin{tabular}{lllll}
    \hline
    &
    \multicolumn{1}{c}{\textbf{Overall}}  & 
    \multicolumn{1}{c}{\textbf{Depth 1}}  & 
    \multicolumn{1}{c}{\textbf{Depth 2}}  & 
    \multicolumn{1}{c}{\textbf{Depth 3}}  \\
    \hline
    $\mathcal{E}_s \in \mathcal{G}_s$ directed to seed set & \num{1648} & \num{1648} & -- & --\\
    
    $\mathcal{E}_m \in \mathcal{G}_m$ & \num{773162} & \num{734896} & \num{36487} & \num{1636} \\

    \# with mentions & \num{374907} & \num{359302} & \num{14920} & 567 \\

    \# with mentions \& retweets & \num{1878} & \num{1765} & 113 & 0 \\

    \# with mentions \& replies & 1183 & 1026 & 292 & 284\\
    \hline
    \# $\mathcal{E}_f \in \mathcal{G}_f$  & \num{27017119} & \num{25042892} & \num{1636161} & 0\\
    \hline
  \end{tabular}
  \end{adjustbox}
  \end{center}
  \caption{Basic statistics of the data crawled}
  \label{table:crawl}
\end{table}

\subsubsection{Ground Truth}\label{sub:agreement}
Following a voting scheme we explain here, we aggregate the votes received for each tweet into a consensus score. We take a pessimistic approach to ensure that a single vote is not decisive in the evaluation of a tweet as abusive (e.g., unlike in Brexit affairs). That is, if the aggregated score is between -1 and 1 the message is considered {\em undecided}. The sum of scores will render a tweet as {\em abusive} in the ground truth when \textgreater 1 and for {\em acceptable} when \textless -1 . The final annotated dataset is comprised of \num{14193} labeled messages, out of which \num{9809} are marked as acceptable and \num{2469} as abusive and \num{1912} undecided.

\begin{figure}
  \centering
    \includegraphics[width=\columnwidth]{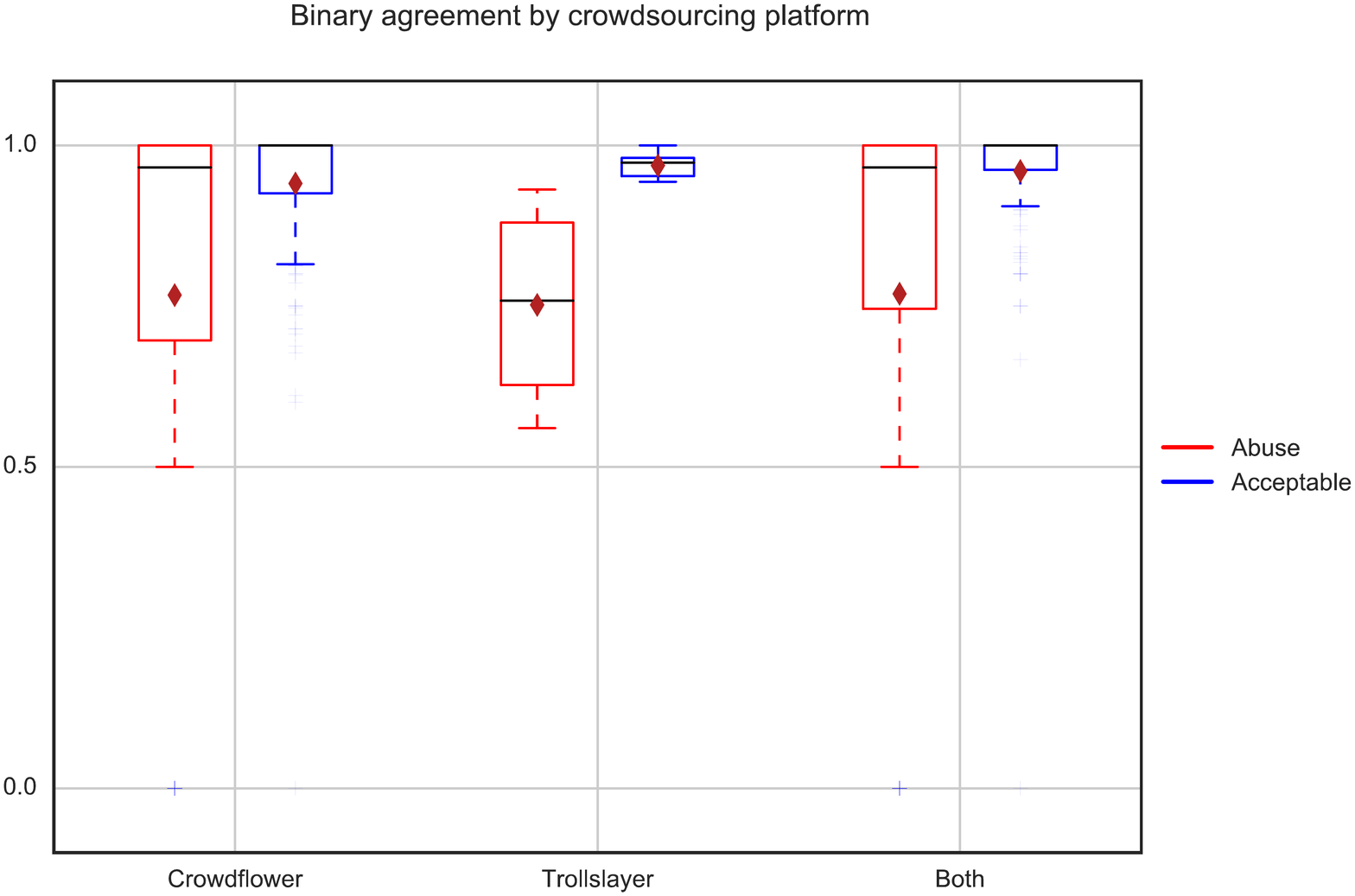}
    \caption{Agreement in ground truth by platform}
    \label{fig:hb-scores}
\end{figure}

Figure~\ref{fig:hb-scores} shows the result of crowdsourcing abuse annotation when asking crowd-workers to mark messages as either, abusive, acceptable or undecided. Agreement is high in both platforms, even so for abusive messages, but as expected lower than acceptable due to perfect disagreement in a number of tweets as the ones we show in Table~\ref{table:disagreement}. There are \num{8} tweets with perfect disagreement in Trollslayer out of around \num{400} annotated, \num{27} in Crowdflower out of \num{720}, and \num{25} in the aggregate out of \num{1912} mentioned above accounting for aggregated voting of all annotations from both platforms. Generally speaking, we see an upper bound of about 3.75\% disagreement for Crowdflower, 2\% in Trollslayer and lower bound of 1.3\% among both, which highlights the importance of employing a minimal set of trusted crowd workers in the annotations (as we did with Trollslayer).

\subsubsection{Agreement}
To ensure agreement among crowd workers is valid, we calculate the inter-assessor agreement score of Randolph's multi-rater kappa~\cite{randolph2005free} among the crowd workers with common tweets annotated. Similarly to Cohen's kappa or Fleiss' Kappa, the Randolph's kappa descriptive statistic is used to measure the nominal inter-rater agreement between two or more raters in collaborative science experiments. We choose Randolph's kappa over the others by following Brennan and Predige suggestion from 1981 of using free-marginal kappa when crowd workers can assign a free number of cases to each category being evaluated (e.g., {\em abusive}, {\em acceptable}) and using fixed-marginal otherwise~\cite{brennan1981coefficient}. Our case considers different crowd workers assigning a different number of annotations to each class or category, which satisfies Randolph's kappa requirement.

Note that in contrast to simple agreement scores, descriptive statistics consider agreement on all three possibilities, {\em abusive}, {\em acceptable} and {\em undecided}, thus providing a more pessimistic measure of agreement among crowd workers. There are number of descriptive statistics~\cite{Warrens2010} such as Light's kappa and Hubert's kappa, which are multi-rater versions of Cohen's kappa. Fleiss' kappa is a multi-rater extension of Scott's pi, whereas Randolph's kappa generalizes Bennett' $S$ to multiple raters.

Given this setting, values of kappa can range from -1.0 to 1.0, with -1.0 meaning a complete disagreement below random, 0.0 meaning agreement equal to chance, and 1.0 indicating perfect agreement above chance. According to Randolph, usually a kappa above 0.60 indicates very good inter-rater agreement. 
Across all annotations we obtain overall agreement of 0.73 and a a Randolph's free-marginal of 0.59 which is about the recommended value in Randolph's kappa (0.60).

\begin{table*}
\centering
\begin{tabular}{|c|c|}
  \hline
  Time & Text \\ \cline{1-2}
  2015-11-26 20:51:49 &
  \multicolumn{1}{m{12cm}|}{RT @Lo100La: @VABVOX @CoralieAlison @caitlin\_roper @MelTankardReist @MelLiszewski If you stand and wait somewhere, men just come and grab you} \\ \cline{1-2}
  2015-11-23 20:41:52 &
  \multicolumn{1}{m{12cm}|}{Yes! Do not submit to the temptation, unless you want to experience an ice pick being driven slowly into your brain @enbrown @MattWelch} \\ \cline{1-2}
  2015-11-29 11:59:25 &
  \multicolumn{1}{m{12cm}|}{@reynardvi @BasimaFaysal @avinashk1975 @SamBamDamdaMan People will literally kill to prove themselves "virtuous". Kill themselves too.} \\ \cline{1-2}
\end{tabular}
\caption{Tweets with perfect disagreement in Both, Trollslayer and Crowdflower}\label{table:disagreement}
\end{table*}

We inspect some of the annotations manually and discover that some scores are aggregated as undecided and not as abusive due to their crowd-workers annotating as undecided several of these tweets serially. That shows the cognitive difficulty in the task of annotating abuse or the tedious nature which we mention before (despite having rewarded the crowd-workers in both platforms). On the other hand, we noticed it is easy for crowd workers to spot offensive messages containing {\em hate speech} or similar (which in fact is abuse but only a subset according to the {\em JTRIG} definition) but not so for deceitful messages or content.



\section{Characterization of Abuse}
This section shows that our method can indeed capture all type of abusive behavior in Twitter and that while humans still have a hard time identifying as abuse deceitful activity, our latest findings suggest the use of network level features to identify some abuse automatically instead.

\subsection{Incidents}
In several cases we find where there is perfect disagreement among crowd workers, see Table~\ref{table:disagreement}; while in others some of the actual abusive ``birds'' are just too difficult to spot for humans given just a tweet but more likely if we inspect an exhaustive list of similar messages from the potential perpetrators' timeline as shown in Table~\ref{table:deceitful}. In that case the abusive ``bird'' is repeatedly mentioning the same users through the special character ``@'' that Twitter enables in order to direct public messages to other participants. Besides, he repeatedly adds a link to a doubtful fund-raising campaign.

\begin{table*}[ht]
\centering
\begin{tabular}{|c|c|c|c|}
  \hline
  Time & Text & Mentions & Hashtags \\ \cline{1-4}
%
  2015-12-11 23:16:25 &
  \multicolumn{1}{m{7cm}|}{TY4follow @CPCharter @EJGirlPolitico @Daboys75! My \#socialentrepreneur \#socialenterprise \#socent @ https://t.co/SgTj5PXJ7H What do U do?} &
  \multicolumn{1}{m{2.5cm}|}{@CPCharter @EJGirlPolitico @Daboys75} &
  \multicolumn{1}{m{2.8cm}|}{\#socialentrepreneur \#socialenterprise \#socent}
\\ \cline{1-4}
  2015-12-11 23:16:27 &
  \multicolumn{1}{m{7cm}|}{TY4follow @CPCharter @EJGirlPolitico @Daboys75! My \#socialentrepreneur \#socialenterprise \#socent @... https://t.co/3t1Kepp8Q5} &
  \multicolumn{1}{m{2.5cm}|}{@CPCharter @EJGirlPolitico @Daboys75} &
  \multicolumn{1}{m{2.8cm}|}{\#socialentrepreneur \#socialenterprise \#socent}
\\ \cline{1-4}
%
%
%
\end{tabular}
\caption{Timeline of potential abusive ``bird'' with deceitful tweets that are not labeled as abuse by humans}\label{table:deceitful}
\end{table*}

We investigate the owner of the Twitter public profile \texttt{@jrbny}: titled ``Food Service 4 Rochester Schools'', which is also related to a presumed founder \texttt{@JohnLester} and both belonging to ``Global Social Entrepreneurship''. 

Firstly, we look into the JSON data of the tweet and check the value of the field {\em source} in the Twitter API just to confirm that it points to ``https://unfollowers.com'', which in turn redirects to ``https://statusbrew.com/'', a commercial site to engage online audiences through social media campaigns. This confirms our suspicions about the nature of the profile and its use for a public fundraising campaign. After a quick inspection at the products offered by this social media campaign management site, indeed we see that the site offers an option to automatically ``schedule content'' for publishing tweets online. In summary, this Twitter account is controlled by humans but uses an automatic scheduling service to post tweets and presumably follow/unfollow other accounts in the hope of obtaining financial donations through an online website. Secondly, expanding the shortened URL linked to tweets as the ones from Table~\ref{table:deceitful}, we find out that indeed the user is redirected to a donation website \footnote{Campaign site: \url{www.pureheartsinternational.com}} from this organization. The site is hosted in Ontario and belongs to the Autonomous System AS62679, namely \textit{Shopify, Inc.}, which reportedly serves several domains distributing malware. We also acknowledge the difficulty in automating crowdsourcing and characterization of the type of abuse {\em deceive}.

Finally, in order to highlight the effect of automated campaign management tools as the ones used in the above case, we crawled the same profile again in 2016-01-10 23:02:59, and the account had only 16690 followers compared to the current 36531 as of January 2017, therefore showing a successful use of semi-automated agents on Twitter for fund-raising activities.


\subsection{Features of Abusive Behavior}
In order to characterize abuse we extract and build a set of novel features, categorized as {\em Attribute} or {\em Graph} based, which measure abuse in terms of the {\em Message}, {\em User}, {\em Social} and {\em Similarity}. We apply Extraction, Transformation and Loading (ETL) on the raw data in order to obtain the inputs to each of the features in those subcategories. The most readily available properties from the \emph{tweet} are extracted. Then we also capture a number of raw inputs in the tweet that identify the features for a particular \emph{user}. The next, and more complex subset of features involve {\em Social} graph metadata, which also enables the computation of the novel {\em Similarity} feature subset, namely the Jaccard index ($\mathcal{J}$). Table~\ref{table:features} summarizes the complete set of features we have developed to evaluate abusive behavior in Twitter.

\begin{table*}
\centering
\begin{adjustbox}{max width=0.8\linewidth}
\begin{tabular}{|l|l|l|l|}
\hline
& Metadata & Feature & Description \\
\hline
\multirow{16}{*}{\rotatebox{90}{\centering \textbf{Attribute based}}}
& \multirow{8}{*}{\rotatebox{0}{\textbf{Message}}}
    & \# mentions & mentions count in tweet\\
    & & \# hashtags & hashtag count in the tweet  \\
    & & \# retweets  & times a message has been reposted \\
    & & is\_retweet (true/false) & message is a repost \\
    & & is\_reply (true/false) & message is a reply \\
    & & sensitive & message links to external URL \\
    & & \#badwords & number of swear words from Google~\cite{googlebadwords} \\
    & & $\nicefrac{\text{\# replies}}{\text{\# tweets} of user}$ & fraction of replies to tweets\\
\cline{2-4}
& \multirow{8}{*}{\rotatebox{0}{\textbf{User}}}
    & verified (true/false) & sender account is verified by Twitter \\
    & & \# favorites & \# tweets marked as favorites by sender \\
    & & age of user account  & days since account creation  \\
    & & \# lists & number of lists of sender\\
    & & $\nicefrac{\text{\# messages}}{\text{age} of user}$ & tweets per day\\
    & & $\nicefrac{\text{\# mentions}}{\text{age} of user}$ & mentions per day \\
    & & $\nicefrac{\text{\# mentions}}{\text{\# tweets} of user}$ & ratio of mentions to tweets\\
    & & account recent & check if account age is $<=$ 30 days\\
\cline{1-4}
\multirow{11}{*}{\rotatebox{90}{\centering \textbf{Graph based}}}
    & \multirow{6}{*}{\rotatebox{0}{\textbf{Social}}}
    & \# subscriptions$^s$ & followee count from public feed of sender \\
    & & \# subscribers$^s$ & follower count to public feed of sender \\
    & & $\nicefrac{\text{\# subscribers}}{\text{age}}$ & ratio of subscribers count to age of sender\\
    & & $\nicefrac{\text{\# subscriptions}}{\text{age}}$ & ratio of subscriptions count to age of sender \\
    & & $\nicefrac{\# \text{subscriptions}}{\# \text{subscribers}}$ & ratio of subscriptions count to subscribers of sender\\
    & & $\nicefrac{\# \text{subscribers}}{\# \text{subscriptions}}$ & ratio of subscribers count to subscriptions of sender\\
    & & reciprocity & true if bi-directional relationship among sender and receiver in $\mathcal{G}_f$\\
    \cline{2-4}
    & \multirow{4}{*}{\rotatebox{0}{\textbf{Similarity}}}
    & $\mathcal{J}$ (subscriptions$^s$, subscriptions$^r$) & $\mathcal{J}$ of sender \& receiver subscriptions\\
    & & $\mathcal{J}$ (subscribers$^s$, subscribers$^r$) & $\mathcal{J}$ of sender \& receiver subscribers\\
    & & $\mathcal{J}$ (subscriptions$^s$, subscribers$^r$) & $\mathcal{J}$ of subscriptions of sender \& subscribers of receiver\\
    & & $\mathcal{J}$ (subscribers$^s$, subscriptions$^r$) & $\mathcal{J}$ of subscribers of sender \& subscriptions of receiver\\
    \hline
\end{tabular}
\end{adjustbox}
\caption{Subsets of features by category}\label{table:features}
\end{table*}

To visualize the data distribution of the most relevant features from Table~\ref{table:features} in detail we show the complementary cumulative distribution function (CCDF), which represents the probability $P$ that a feature having value of $\geq x$ in the x axis does not exceed $X$ in the y axis. We use the CCDF in log-log scale to be able to pack a large range of values within the axis of the plot.

\begin{figure}[!p]
\centering
\subfloat[\#hashtags in tweet]{
 \includegraphics[width=0.25\textwidth]{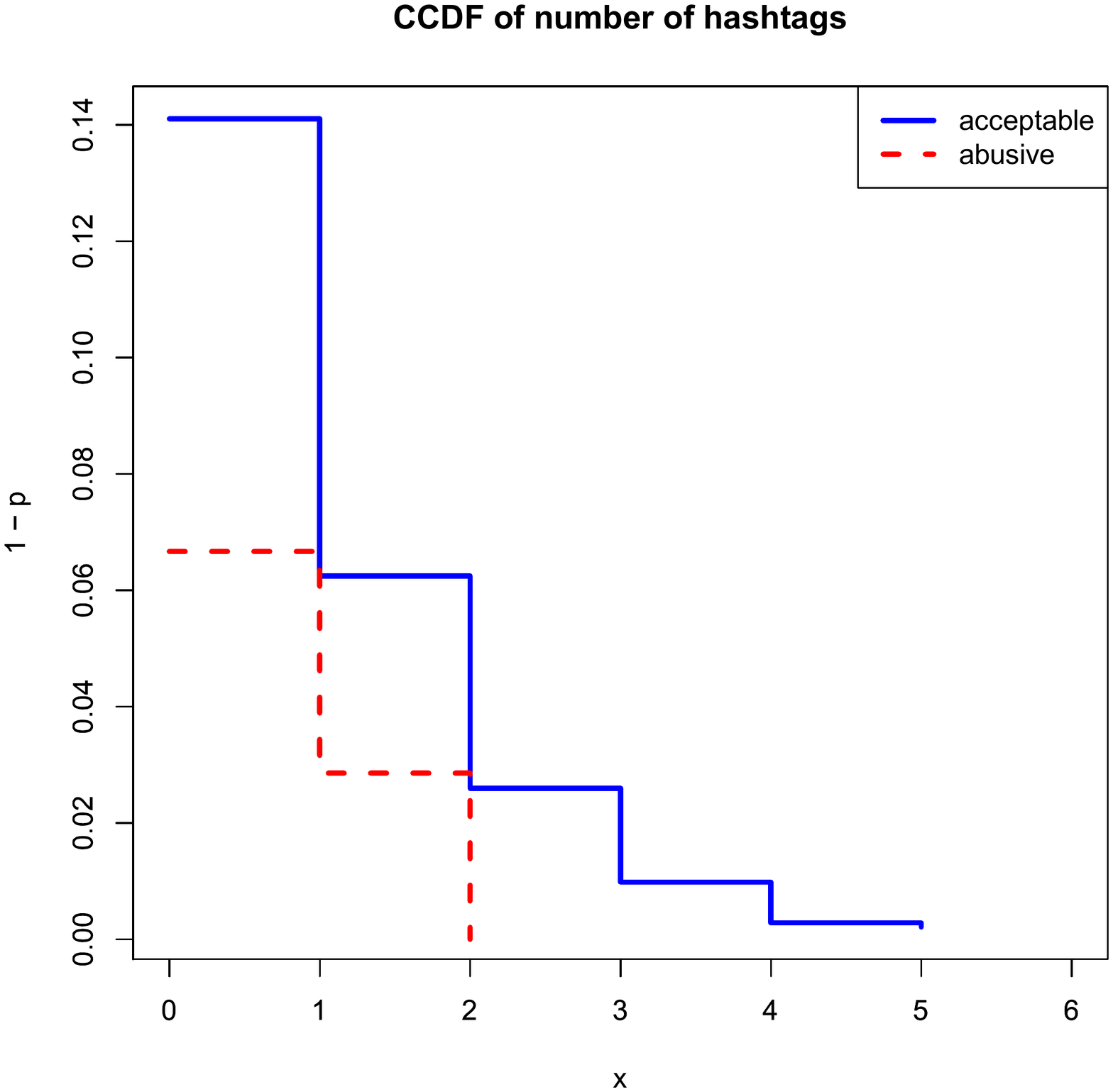}\label{fig:ccdf-hashtags}
}
\subfloat[\#mentions in tweet]{
 \includegraphics[width=0.25\textwidth]{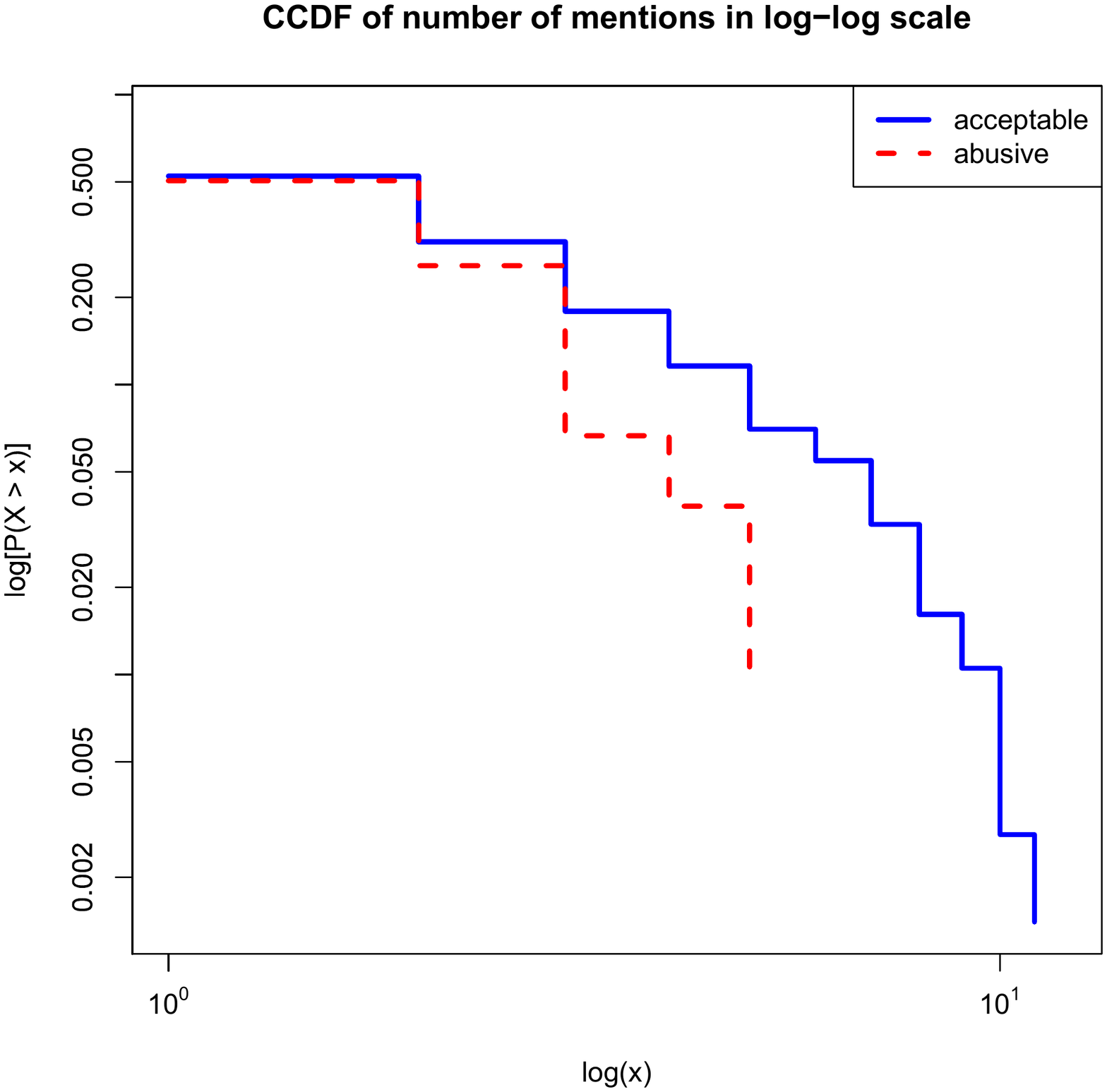}\label{fig:ccdf-mentions}
}

\subfloat[\#badwords in tweet]{
  \includegraphics[width=0.25\textwidth]{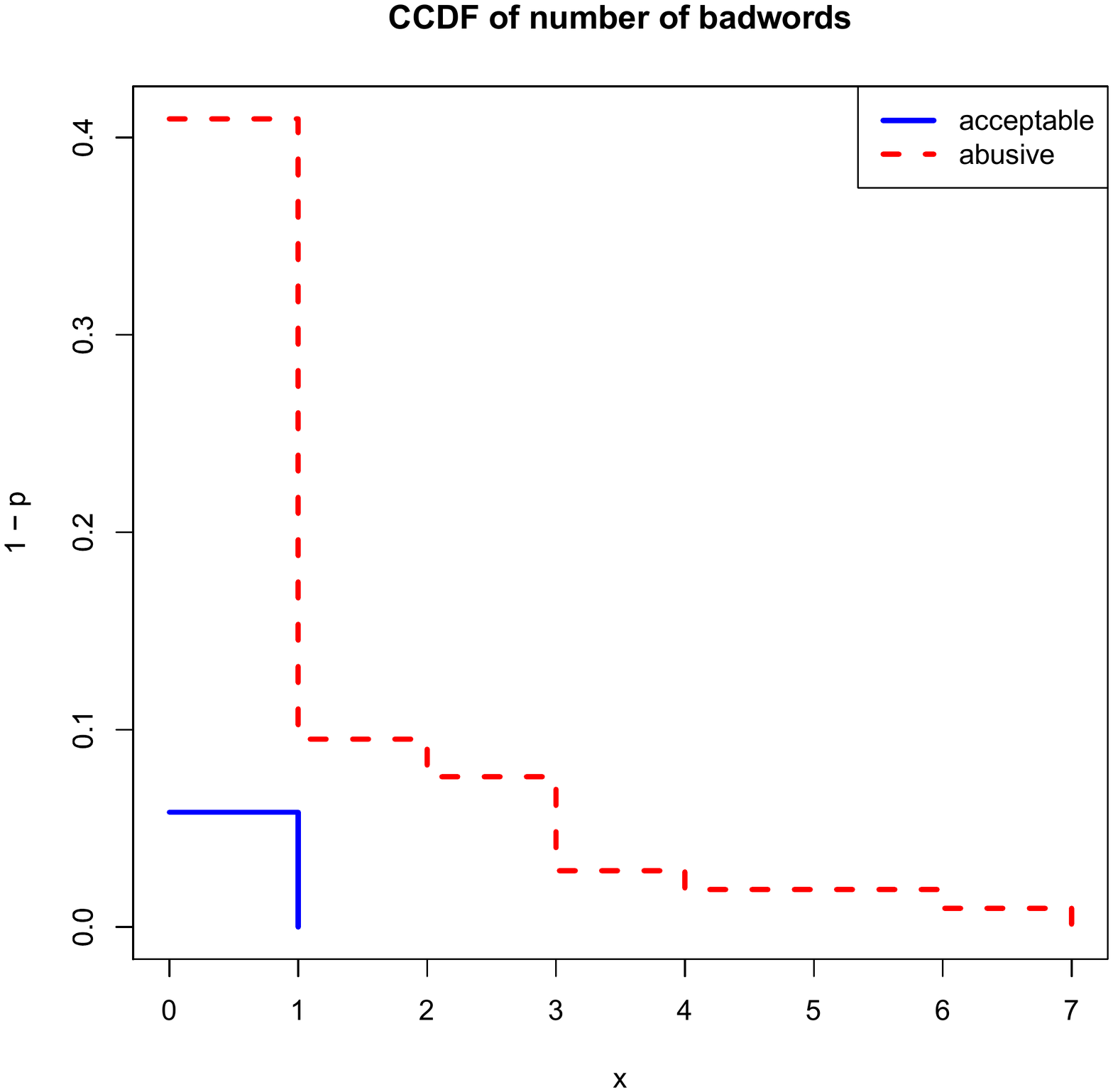}\label{fig:ccdf-badwords}
}
\subfloat[\#Replies over tweets]{
  \includegraphics[width=0.25\textwidth]{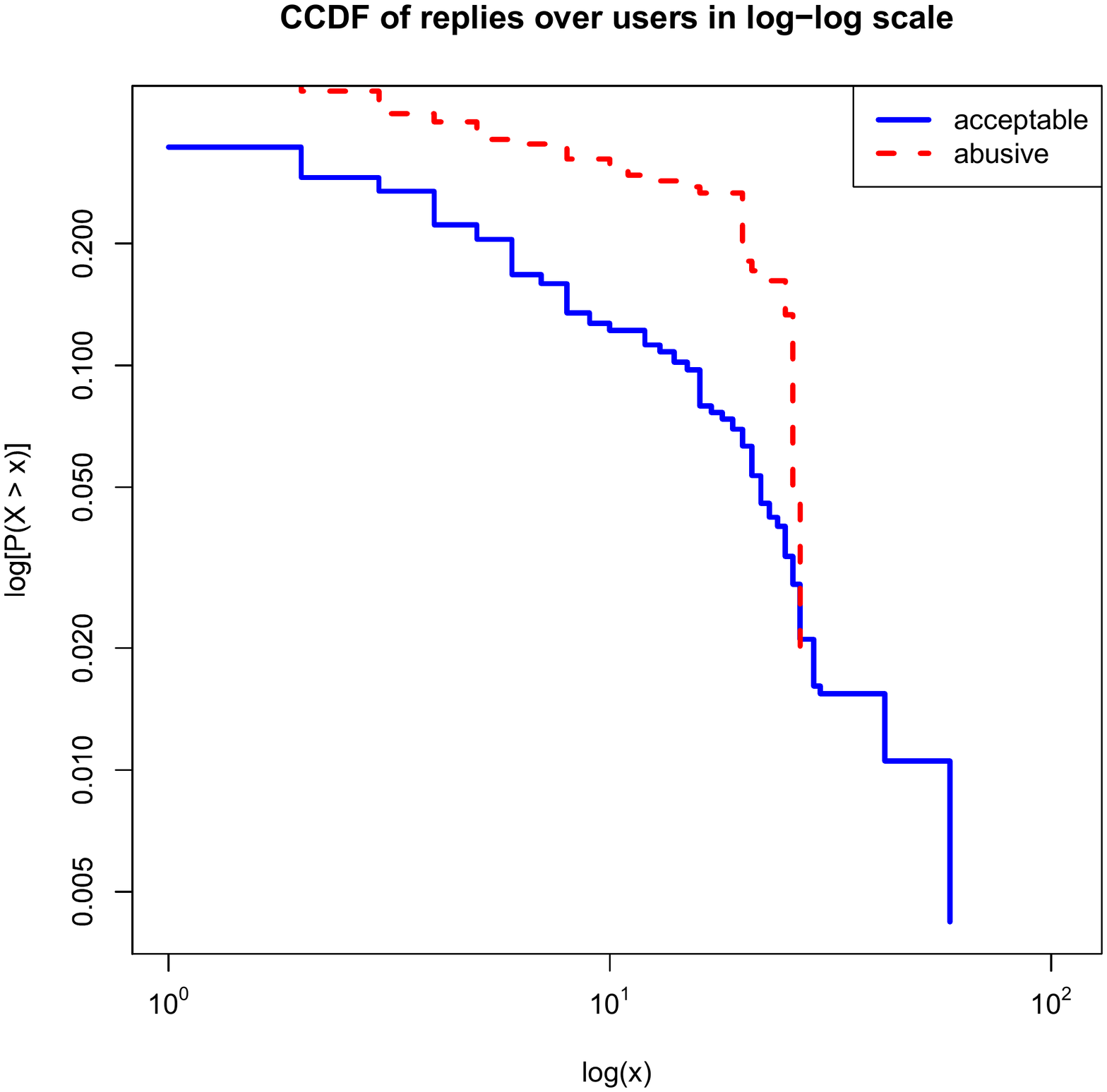}\label{fig:ccdf-replies}
}

\subfloat[\#favorites]{
 \includegraphics[width=0.25\textwidth]{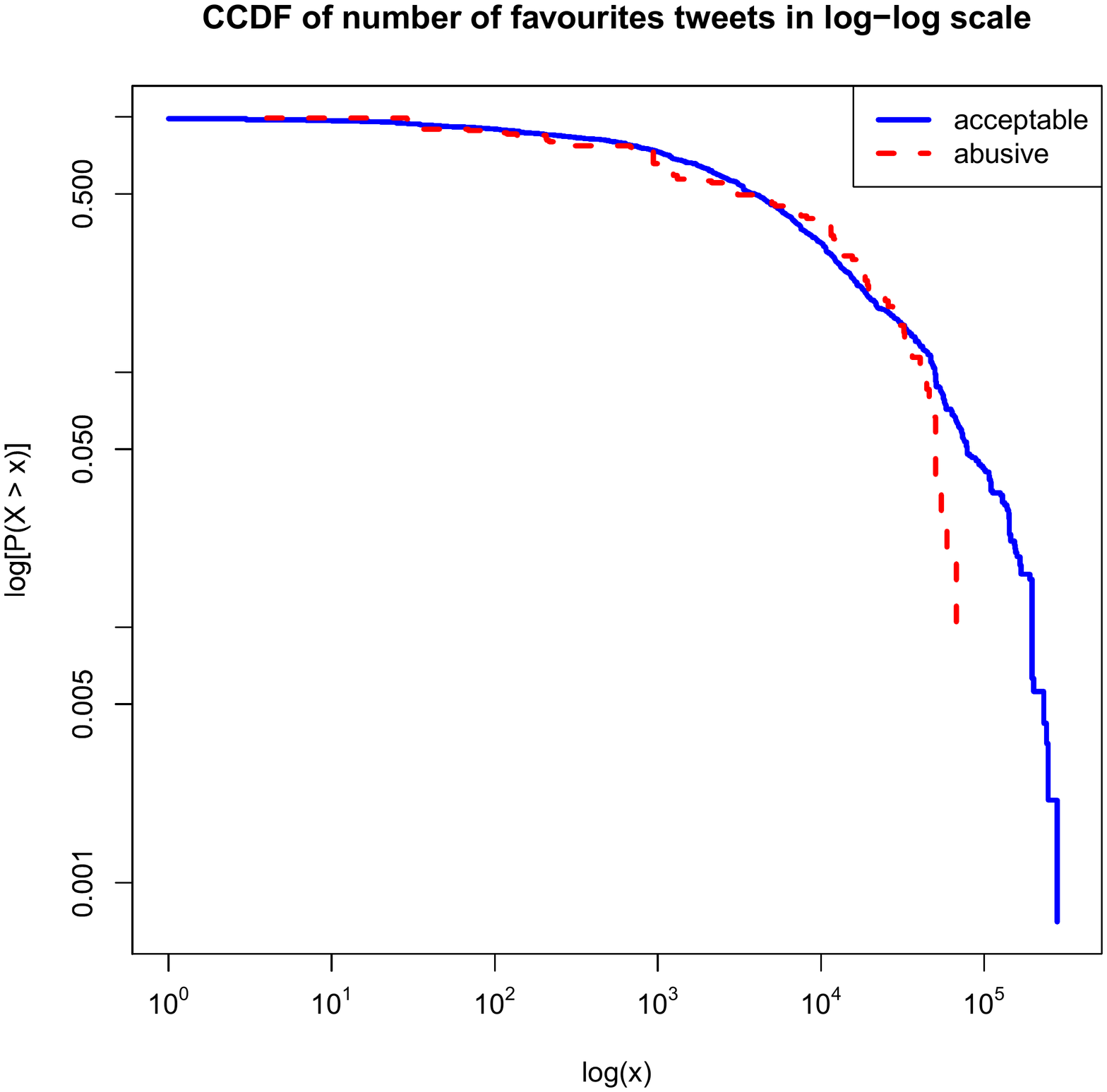}\label{fig:ccdf-favorites}
}
\subfloat[Age of user account]{
  \includegraphics[width=0.25\textwidth]{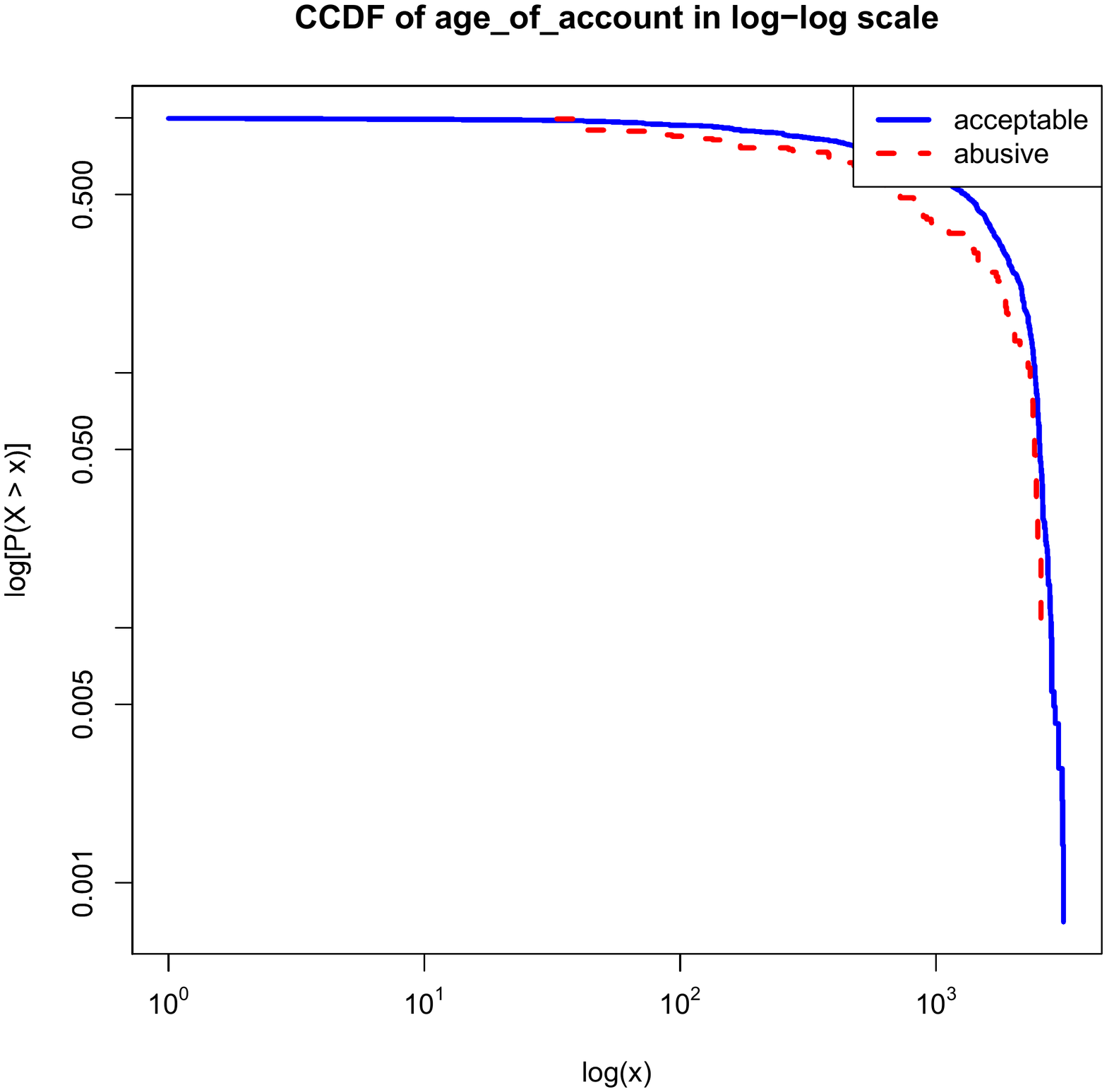}\label{fig:ccdf-age-of-account}
}

\subfloat[\# of tweets per day]{
  \includegraphics[width=0.25\textwidth]{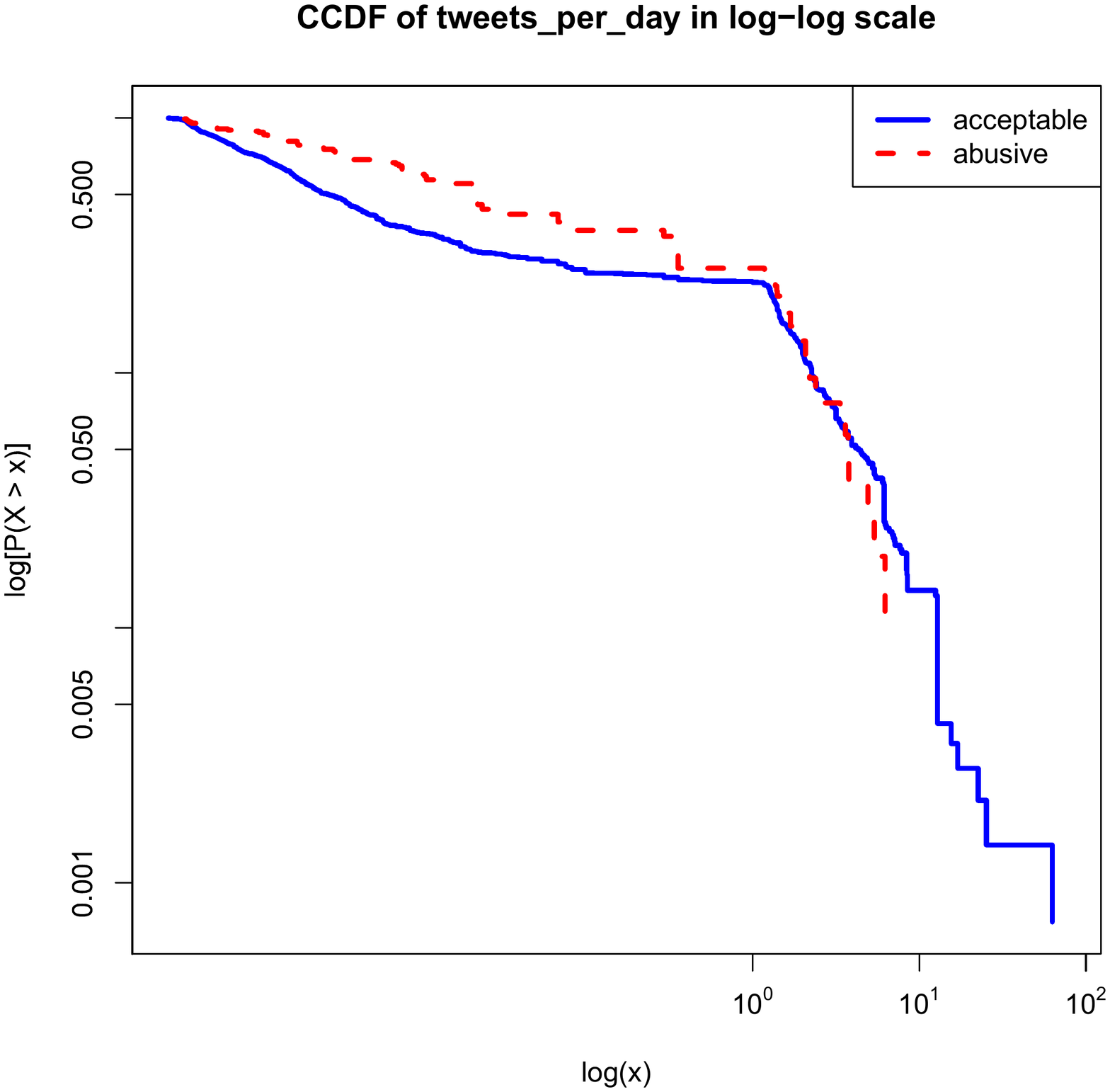}\label{fig:ccdf-tweets-day}
}
\subfloat[\#user lists]{
 \includegraphics[width=0.25\textwidth]{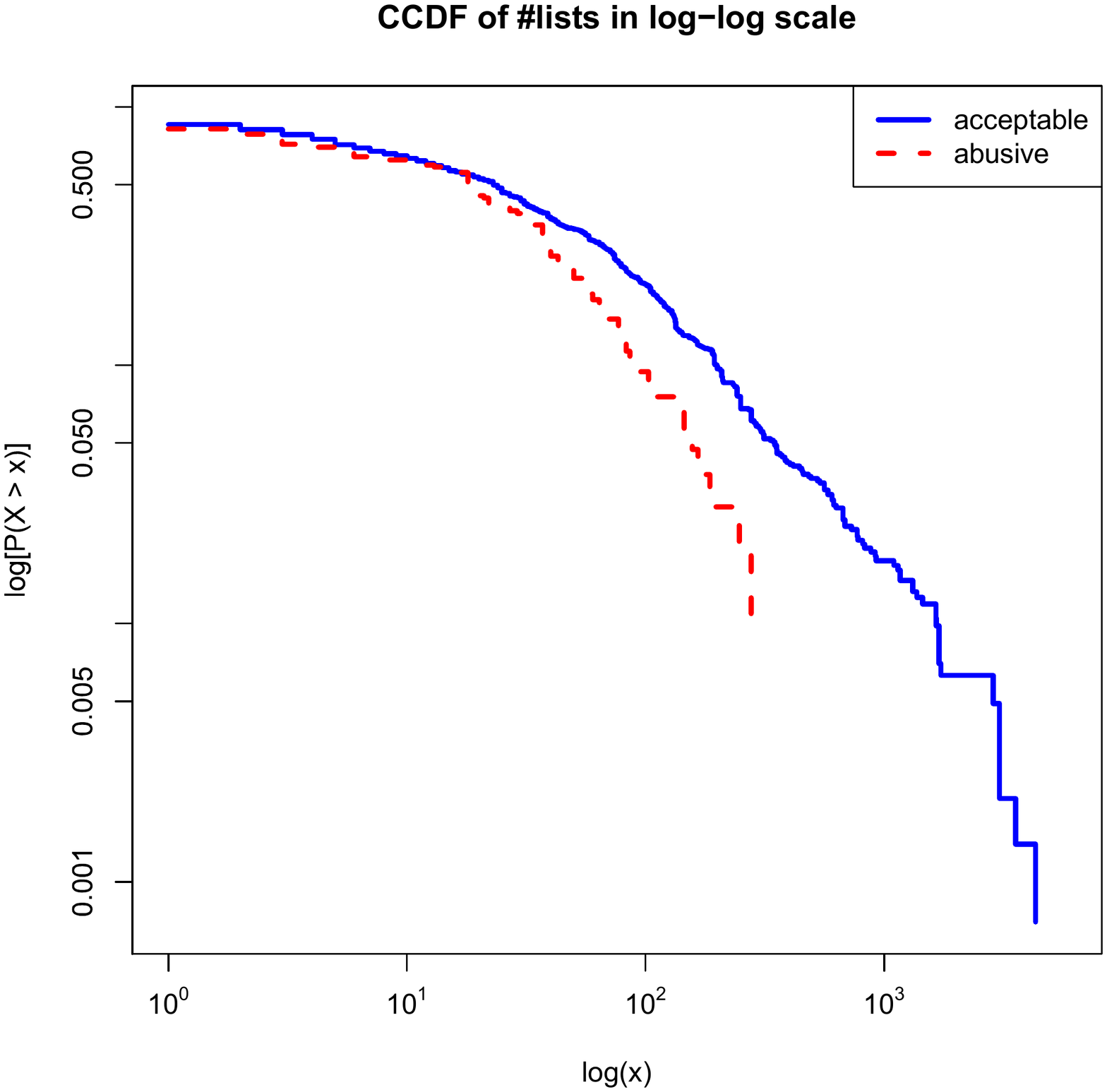}\label{fig:ccdf-account-lists}
}

\subfloat[\# of retweets]{
  \includegraphics[width=0.25\textwidth]{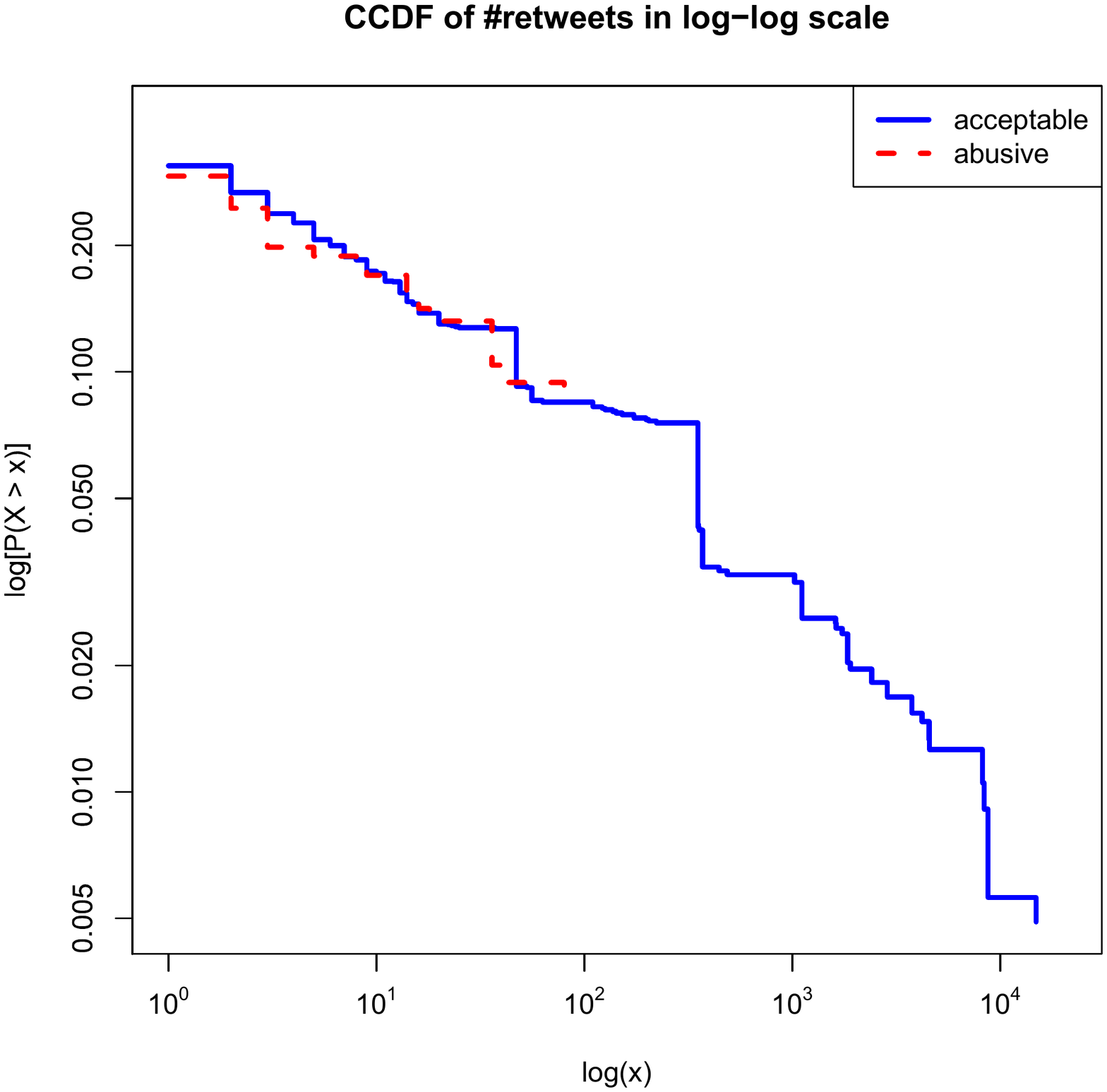}\label{fig:ccdf-retweet-count}
}
\caption{Attribute based features}
\label{fig:attributes}

\end{figure}
\begin{figure}
\centering
\subfloat[\#Subsriptions]{
 \includegraphics[width=0.25\textwidth]{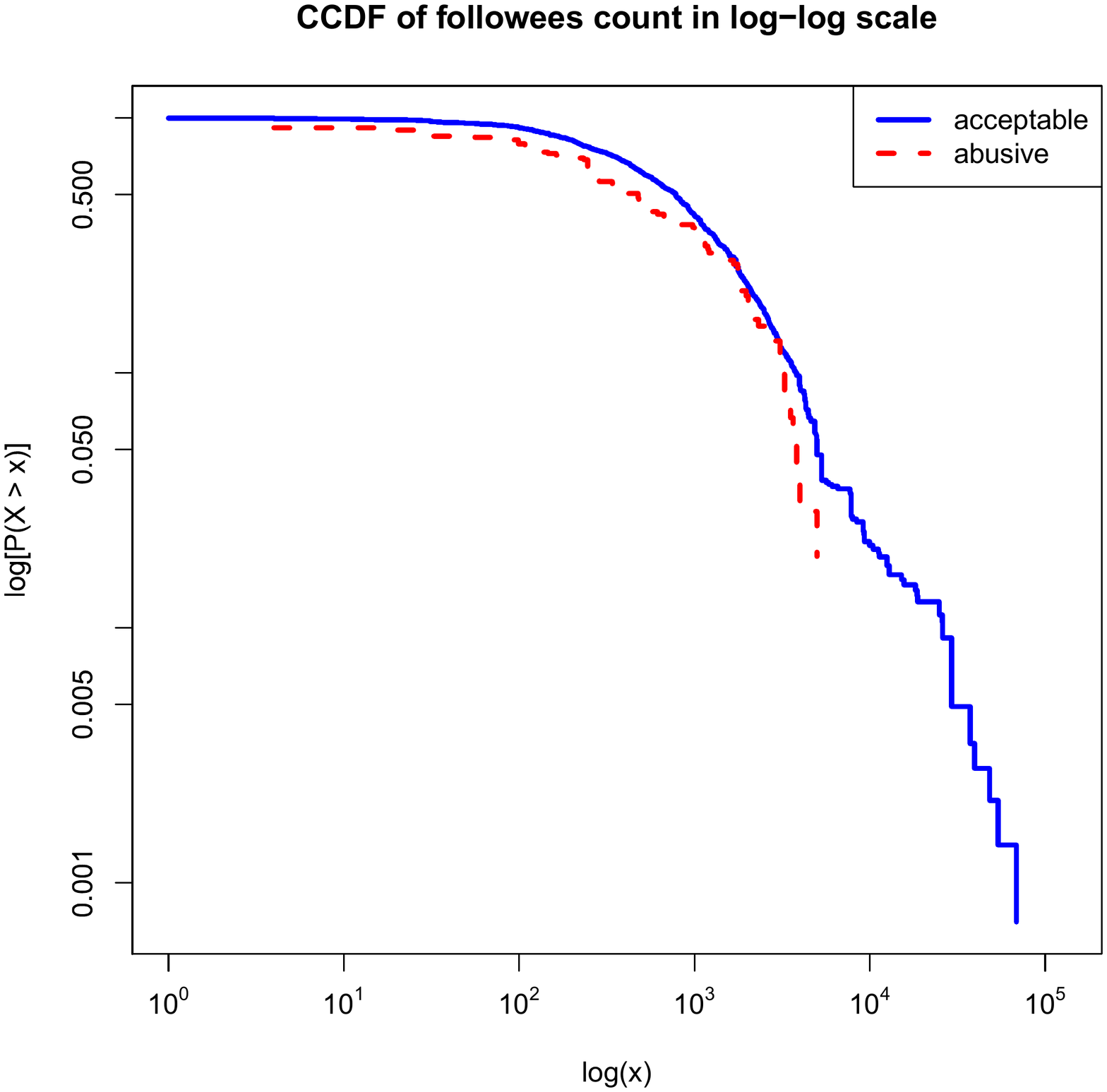}\label{fig:ccdf-followees}
}
\subfloat[\#Subscribers]{
 \includegraphics[width=0.25\textwidth]{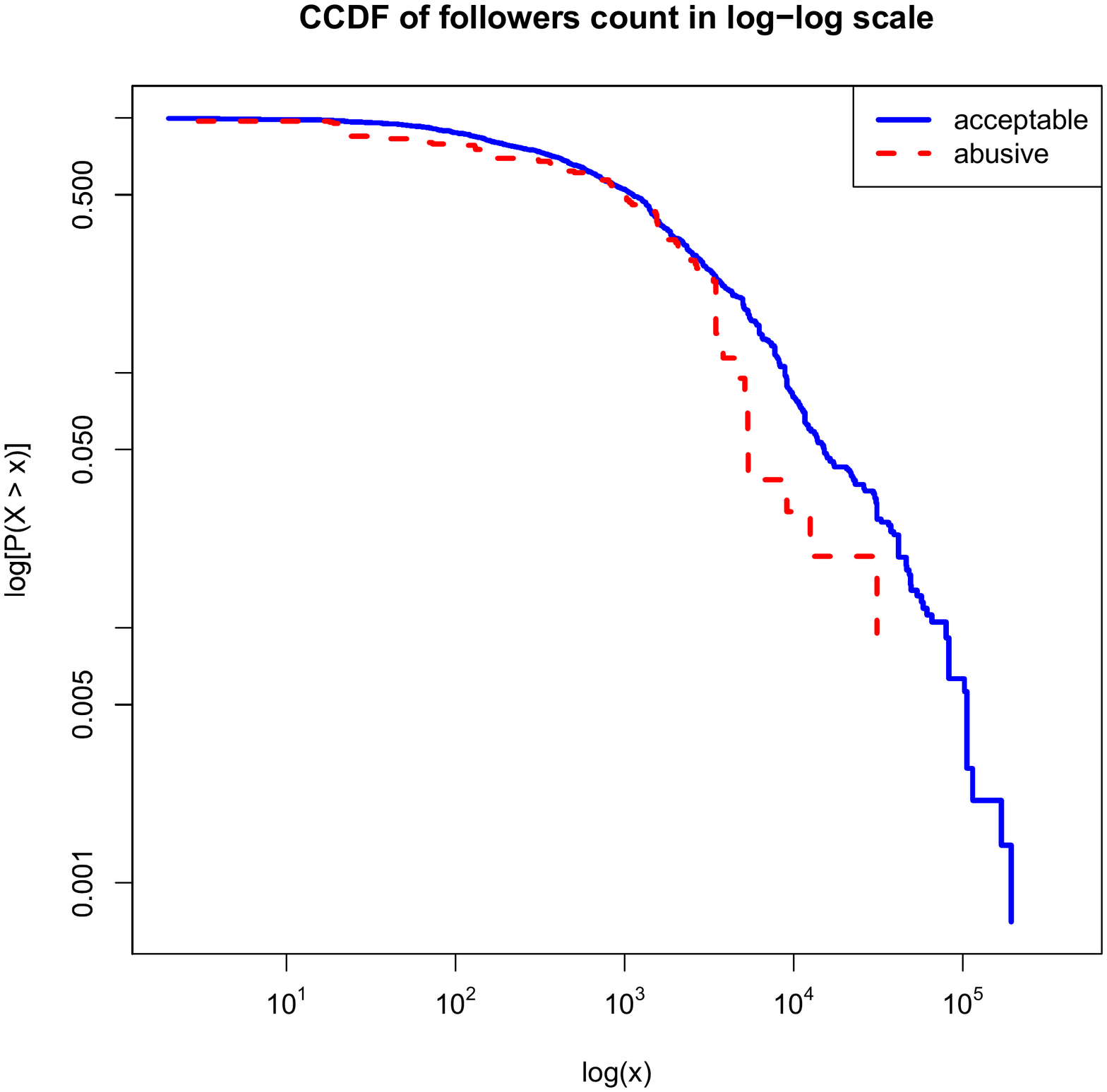}\label{fig:ccdf-followers}
}

\subfloat[\#Subscribers per day]{
  \includegraphics[width=0.25\textwidth]{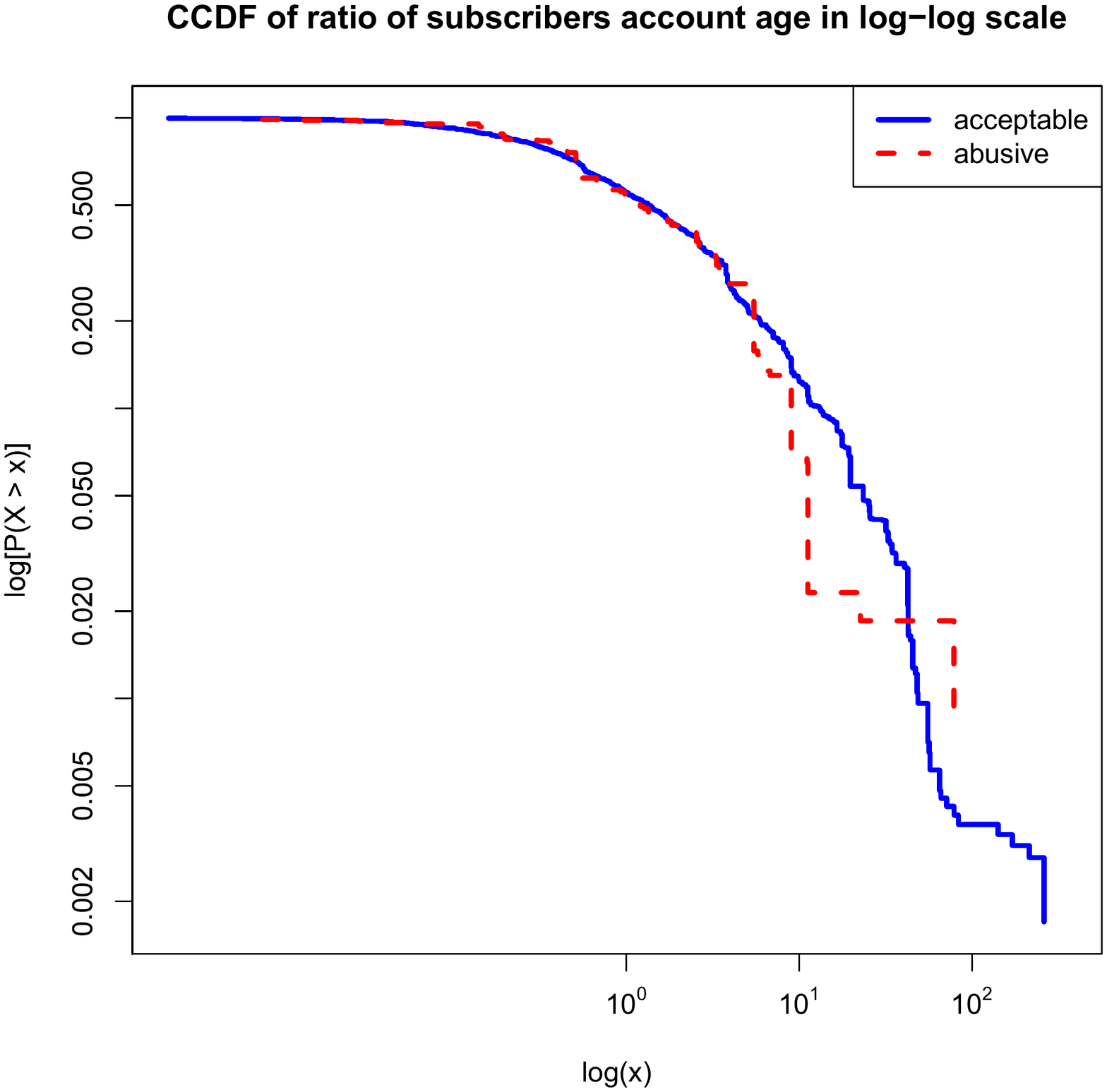}\label{fig:ccdf-ratio-follows-received}
}
\subfloat[\#Subscriptions per day]{
  \includegraphics[width=0.25\textwidth]{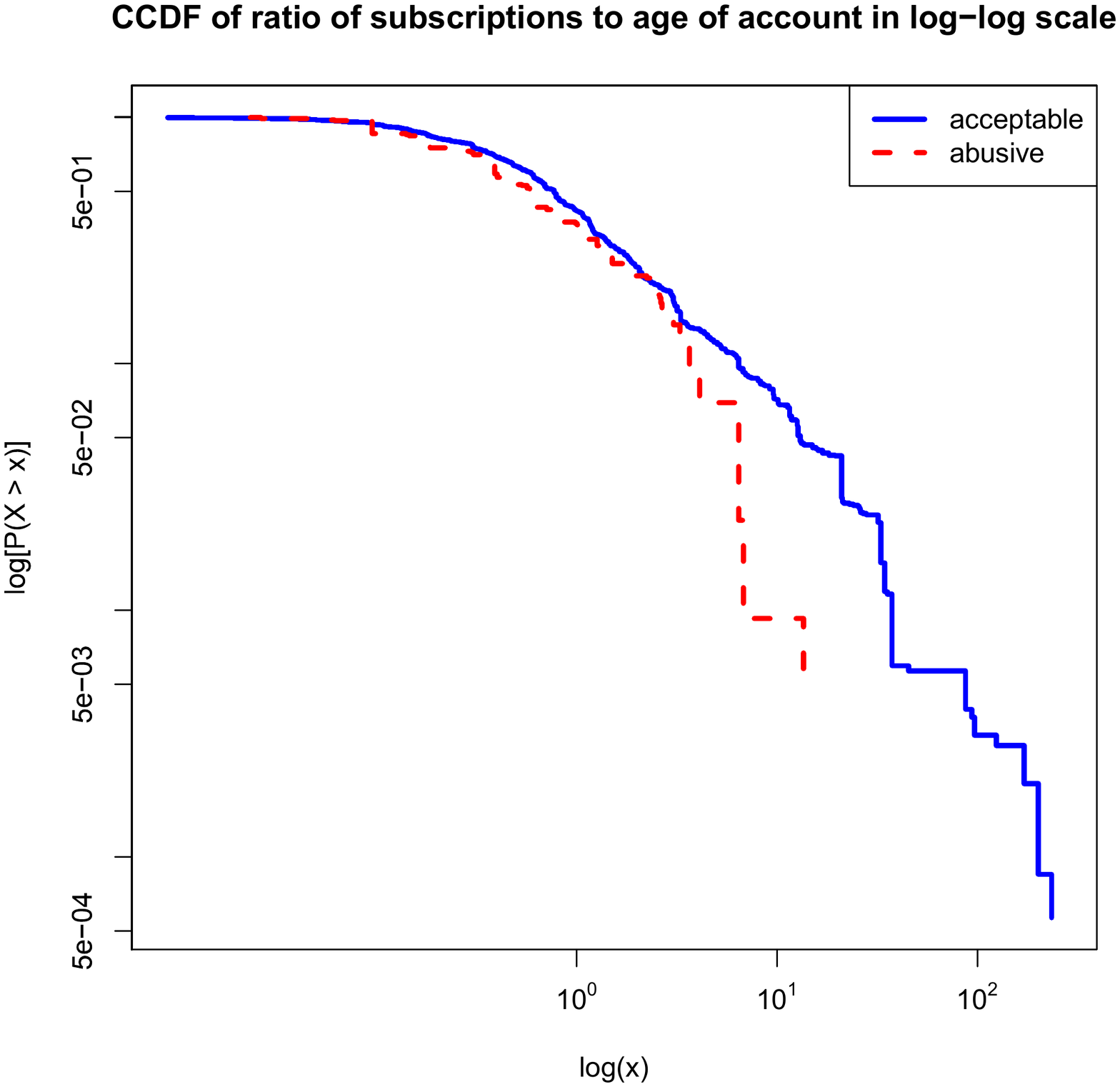}\label{fig:ccdf-ratio-follows-sent}
}

\subfloat[$\frac{\#Subscribers}{\#Subscriptions}$]{
  \includegraphics[width=0.25\textwidth]{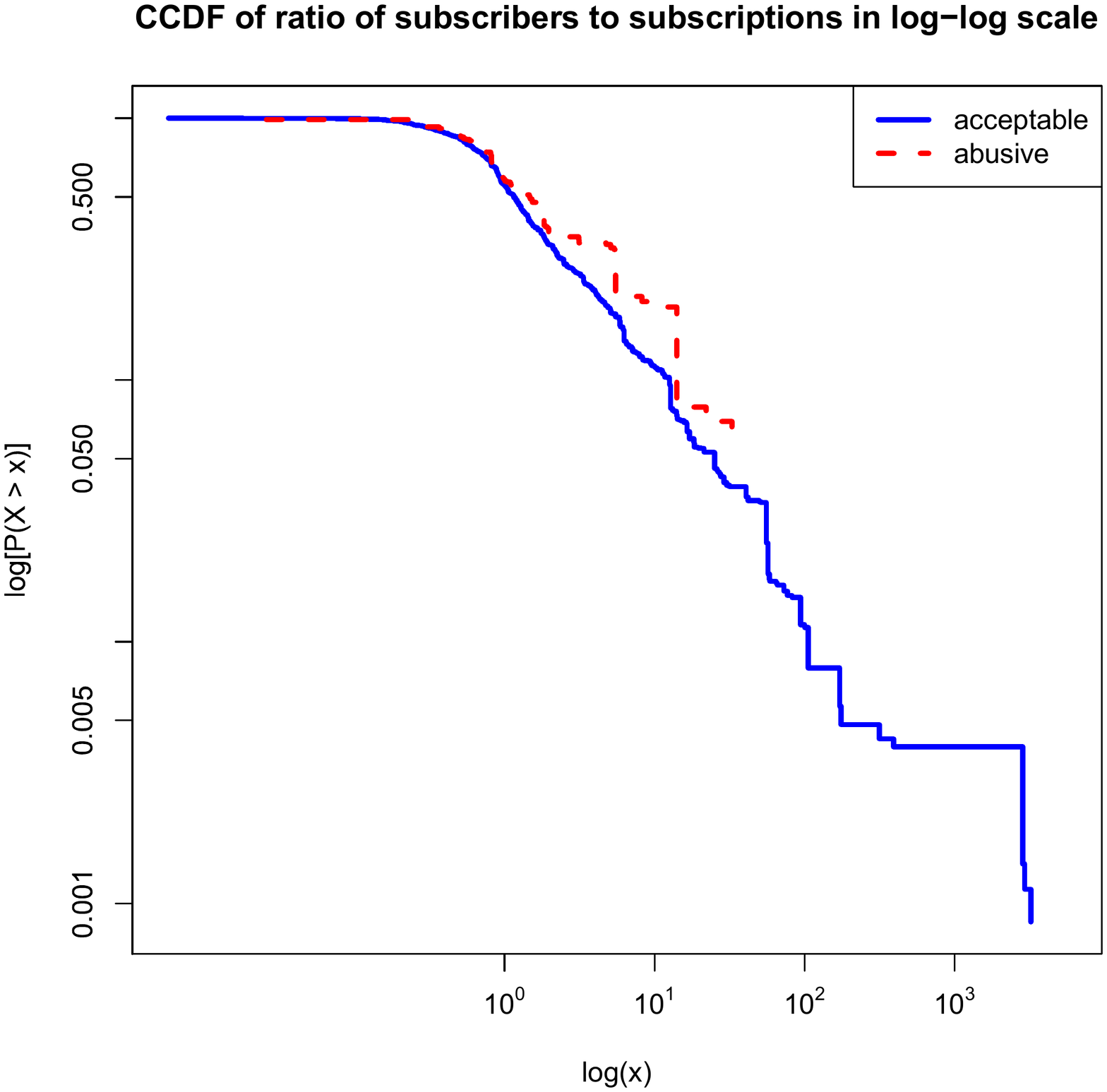}\label{fig:ccdf-ratio-followers-followees}
}
\subfloat[$\frac{\#Subscriptions}{\#Subscribers}$]{
  \includegraphics[width=0.25\textwidth]{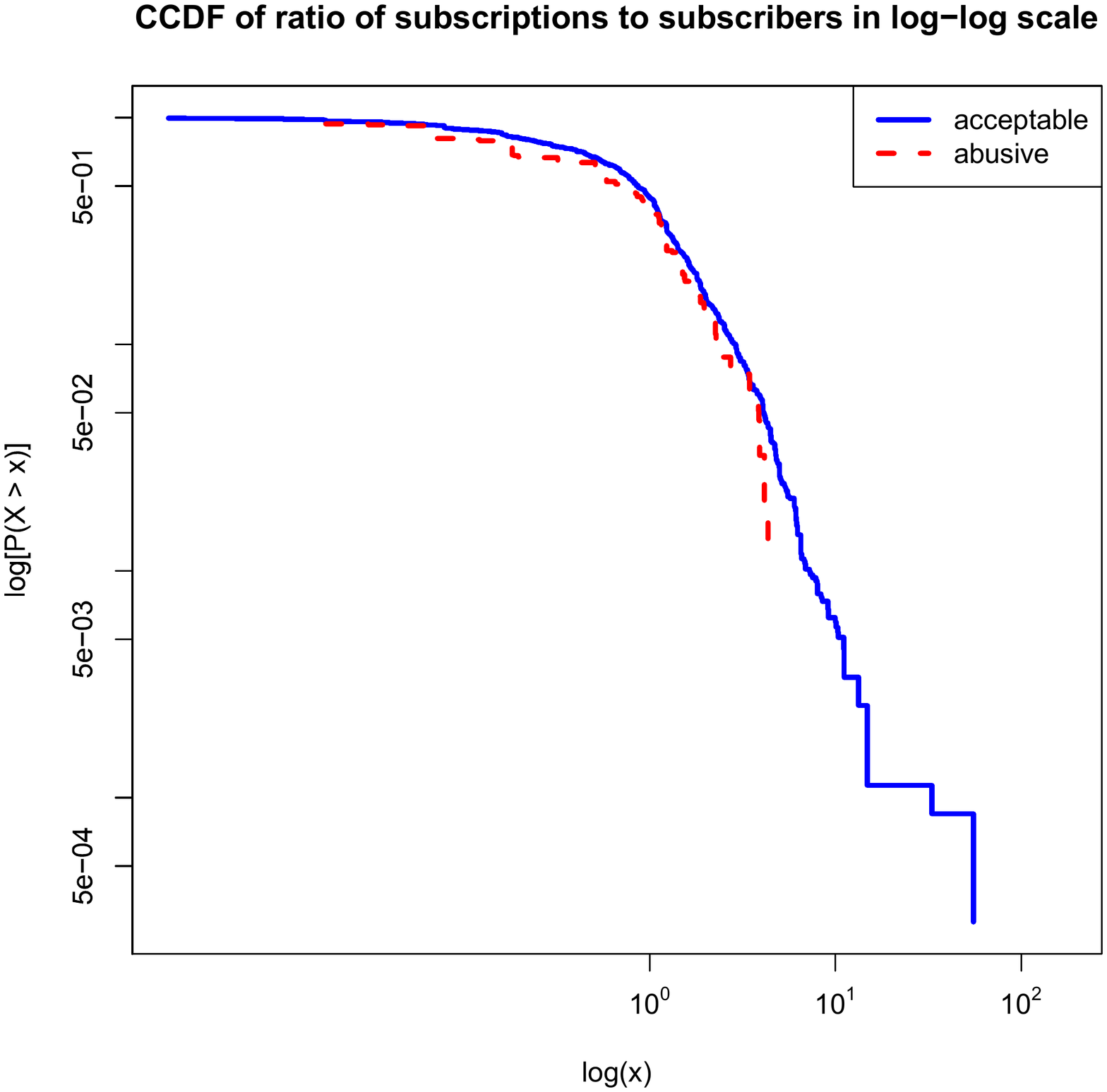}\label{fig:ccdf-ratio-followees-followers}
}

\subfloat[$\mathcal{J}\frac{subscription^s}{subscription^r}$]{
 \includegraphics[width=0.25\textwidth]{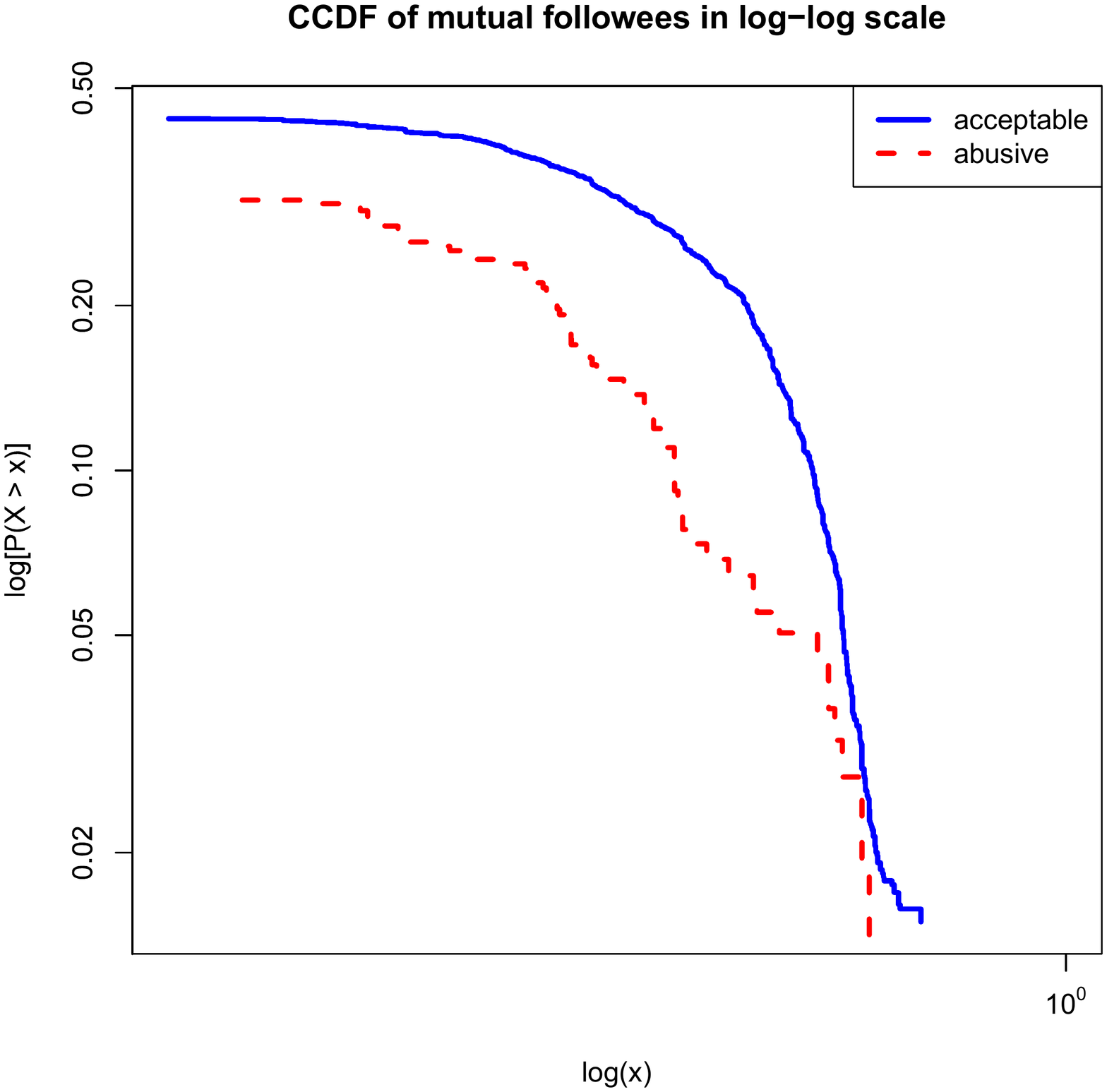}\label{fig:ccdf-followees-followees}
}
\subfloat[$\mathcal{J}\frac{subscriber^s}{subscriber^r}$]{
 \includegraphics[width=0.25\textwidth]{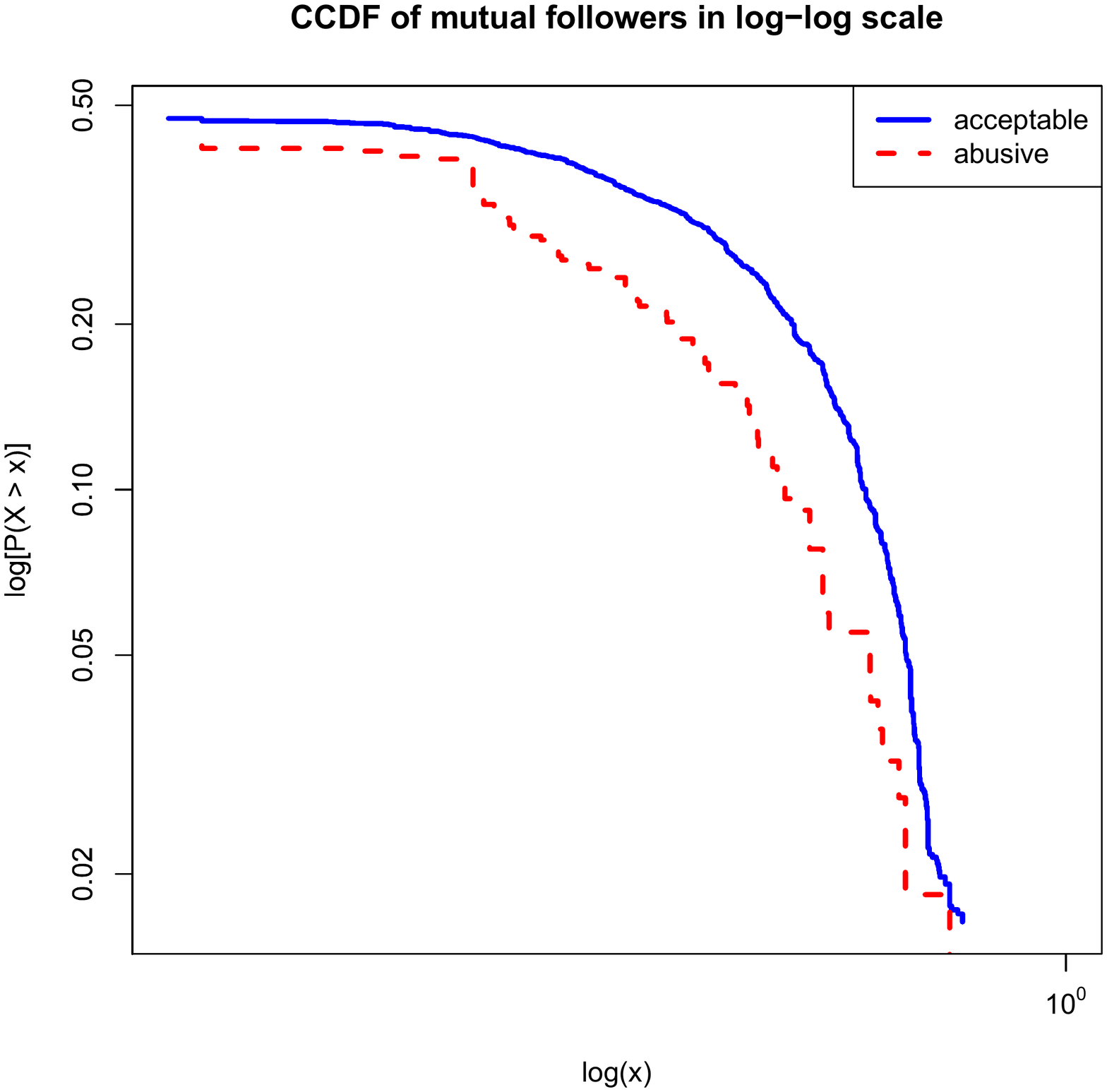}\label{fig:ccdf-followers-followers}
}

\subfloat[$\mathcal{J}\frac{subscriber^s}{subscription^r}$]{
  \includegraphics[width=0.25\textwidth]{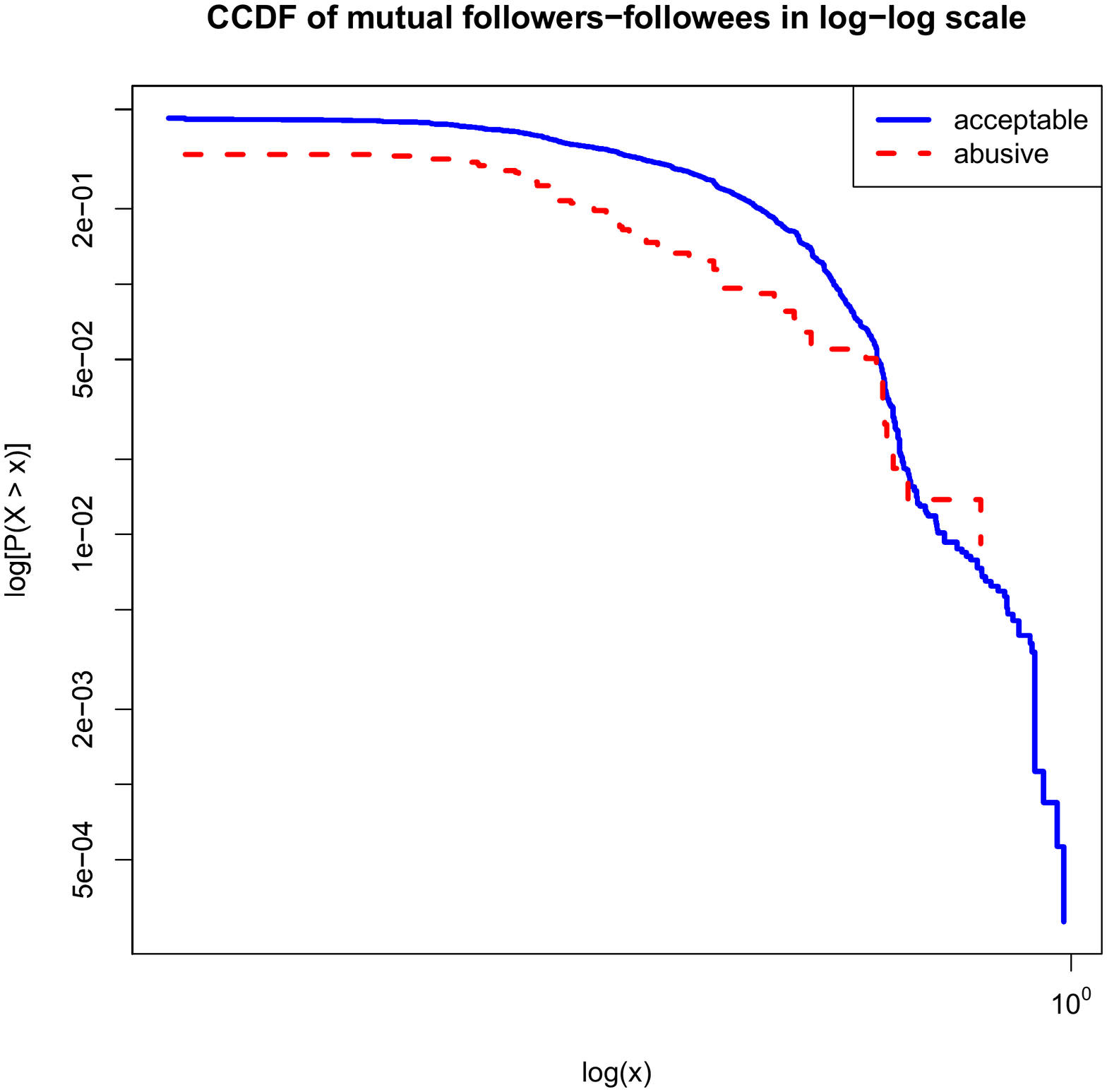}\label{fig:ccdf-followers-followees}
}
\subfloat[$\mathcal{J}\frac{subscription^s}{subscriber^r}$]{
  \includegraphics[width=0.25\textwidth]{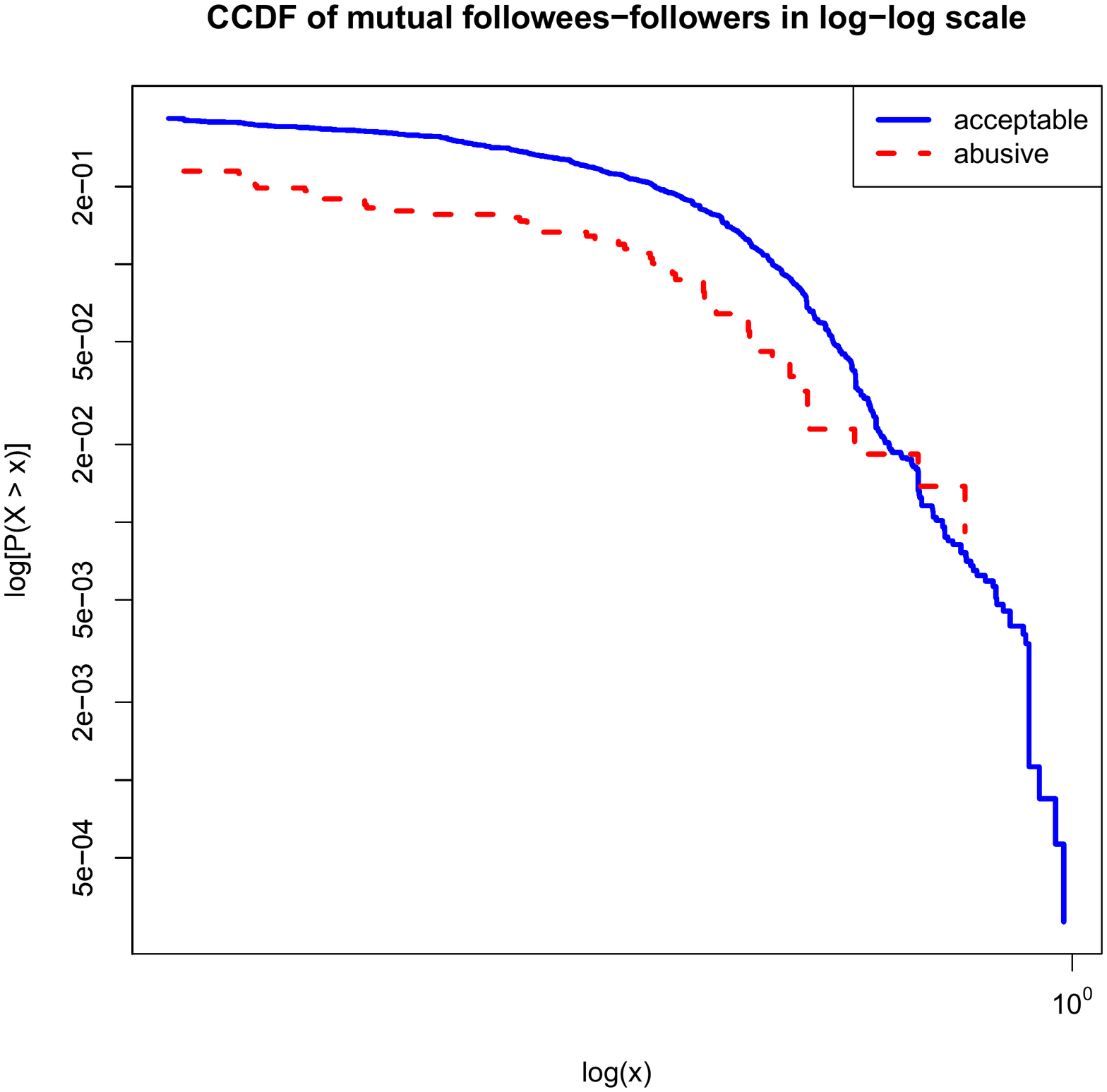}\label{fig:ccdf-followees-followers}
}
\caption{Graph-based features}
\label{fig:social-and-similarity}
\end{figure}

In~\Cref{fig:attributes,fig:social-and-similarity} we compare the characteristic distribution among abuse and acceptable content in our annotated dataset. The dotted line here represents abusive while the continuous one acceptable. 

For the {\em Attribute} based features we notice the most significant gap among acceptable and abusive is the {\em Message} category, in particular the number of replies that a sender user has authored, meaning that abusive ``birds'' reply more often and seek controversy as part of their public speech in Twitter. This makes sense from a ``trolling'' perspective if we consider that the definition of troll is a user that posts controversial, divisive and at times inflammatory content. Secondly, and to the contrary of what we expected, we observe that humans agree on abuse when there are fewer receivers or mentioned users, so the abuse is less likely to be directed to multiple victims according to this. Otherwise, Table~\ref{table:disagreement} shows that no agreement is reached with multiple targets if addressing users as a group, which can not be correlated into a personal attack to the potential victim. We see this as an indication of perpetrators sending disguising messages to their victims in order to decrease the visibility of their abusive behavior.

Finally, the distribution presented in the ``badwords'' feature shows that at least one ``badword'' exist for many of tweets annotated as abusive by our crowd workers, showing a light tailed distribution with smaller probabilities for a larger number of ``badwords''. Firstly, this confirms that human crowd workers are notably good at flagging abusive content when it is related to the language itself and secondly, that abusive messages flagged as such by humans did not contain many ``badwords''. That is also confirmed by the fact that ``bad words'' have a negligible value in the distribution of acceptable for such feature. On the contrary, with hashtags we mostly observe acceptable messages in the CCDF thus indicating that messages from our ground truth flagged as abusive barely contain any hashtags.


We observe that some of the similarity features in the {\em graph-based} category exhibit a distinguishable pattern among acceptable and abusive messages. In particular, this is the case for {\em mutual subscribers} and {\em mutual subscriptions}, where the feature is calculated using {\em Social} graph metadata from a pair of users, namely sender and receiver. The most interesting CCDF is perhaps the {\em mutual subscriptions} one, Figure~\ref{fig:ccdf-followees-followees}, in which there is a significant initial gap between the social graph of acceptable and abusive messages in the log probability ($P(X>x)$) in the y axis for nearly about two-thirds of the distribution. Note that here the maximum value of the axis runs from zero to $10^0$ given that we compute similarity using Jaccard. Considering that we did not present crowd workers with information about the social graph, it is quite surprising that some of these the graph-based features show a characteristic pattern.
\section{Related Work}
The following section covers works similar to ours that fall in the categories of the included subsections.
\subsection{Graph-based}
To characterize abuse without considering the content of the communication, graph-based techniques have been proven useful for detecting and combating dishonest behavior~\cite{Ortega2013} and cyberbullying~\cite{Galan-Garcia2014}, as well as to detect fake accounts in OSN~\cite{Cao2012}. However, they suffer from the fact that real-world social graphs do not always conform to the key assumptions made about the system. Thus, it is not easy to prevent attackers from infiltrating the OSN or micro-blogging platform in order to deceive others into befriending them. Consequently, these Sybil accounts can still create the illusion of being strongly connected to a cluster of legitimate user accounts, which in turn would render such graph-based Sybil defenses useless. On the other hand and yet in the context of OSN, graph-based Sybil defenses can benefit from supervised machine learning techniques that consider a wider range of metadata as input into the feature set in order to predict potential victims of abuse~\cite{boshmaf2015thwarting}. 
Facebook Immune System (FIS) uses information from user activity logs to automatically detect and act upon suspicious behaviors in the OSN.  Such automated or semi-automated methods are not perfect. In relation to the FIS, \cite{boshmafbots2011} found that only about 20\% of the deceitful profiles they deployed were actually detected, which shows that such methods result in a significant number of false negatives.

\subsection{Victim-centric}
The data collection in~\cite{garcia2016discouraging} was partially inspired by the idea of analyzing the victims of abuse to eventually aid individual victims in the prevention and prediction of abusive incidents in online forums and micro-blogging sites as Twitter. One observation from previous research~\cite{boshmaf2015integro} that we have embedded into some of our features is that abusive users can only befriend a fraction of real accounts. Therefore, in the case of Twitter that would mean having bidirectional links with legitimate users. We capture that intuition during data collection by scraping in real-time the messages containing mentions to other users ({\tt @user}) and thus we are able to extract features such as ratio of follows sent/received, mutual subscribers/subscriptions, etc.

\subsection{Natural Language Processing and text based}
Firstly, previous datasets in this area are not yet released or in their infancy for verification of their applicability as abuse ground truth gold standard. The authors of~\cite{nobata2016abusive} claim to outperform deep learning techniques to detect hate speech, derogatory language and profanity. They compare their results with a previous dataset from~\cite{Djuric:2015} and assess the accuracy of detecting abusive language with distributional semantic features to find out that it does largely depends upon the evolution of the content that abusers post in the platform or else having to retrain the model.

Finally, it is worth mentioning we in our feature set do not include sentiment analysis inputs as~\cite{slangsd} did; simply because we are interested in complex types of abuse that require more than just textual content analysis. Additionally, we have noticed that while some words or expressions may seem abusive at first (e.g., vulgar language), they are not when the conversation takes place between participants that know each other well or are mutually connected in the social graph (e.g., family relatives).

\subsection{Other datasets}
Following the above classifications, we compile a number of previous works~\cite{de2010does,cha2010measuring,kwak2010twitter,gabielkov2012} that collected a large portion of the Twitter graph for its characterization but not really meant for abusive behavior. Note some of these datasets can provide some utility from their social-graph for characterization of abusive behaviour but they are either anonymized or we are not able to get access to them. Naturally, social-graph metadata is not available due to restrictions imposed by Twitter Terms and Conditions (TTC) for data publishing. We also find the Impermium dataset, from a public Kaggle competition~\cite{impermium-dataset} that provides the text of a number of tweets and labels for classifying such messages as an insult or not. This can be useful for textual analysis of abuse (only for non-subtle insults), which can be supported by application of NLP based techniques, but it does not contain any social graph related metadata that we use in our characterization of abuse. Besides, as the tweet identifiers from the Imperium dataset are anonymized, it is not possible to reproduce data collection.

\section{Conclusion}
We concluded that identifying abuse is a hard cognitive task for crowd workers and that it requires employing specific guidelines to support them. It is also necessary to provide a platform as we created or questionnaires to ask crowd workers to flag a tweet as abusive if it falls within any of the categories of the guidelines, in our case the 4 D's of JTRIG, {\em deny}, {\em disrupt}, {\em degrade}, {\em deceive}. As a crowd worker provides a non-binary input value from {\em acceptable}, {\em abusive}, {\em undecided} to annotate tweets from $\mathcal{E}_m$, the latter option is important; even with relatively clear guidelines, crowd workers are often unsure if a particular tweet is abusive. To further compensate for this uncertainty, each tweet has been annotated multiple times by independent crowd workers (at least 3). We highlight the reason for the disagreement we encountered by listing a few tweets in Table~\ref{table:disagreement}. Table~\ref{table:deceitful} contains metadata from a user that consistently tweets from a third-party tweet scheduling service.

Additionally, using the set of features presented here one could provide semi-automated abuse detection in order to help humans to act as judges of abuse. Filtering ``badwords'' is not quite enough to judge a user as abusive or not, so in order to provide a better context to human crowd workers one could imagine coupling the score of attribute based features with those graph-based features that can provide an implicit nature of the relationships between senders and receivers of the content, thus flagging messages or users as abusive ``bird'' (or not) in Twitter. This will also present an scenario where abuse is a less tedious and self-damaging tasks for human crowd workers reading abusive content during annotation.

\ifCLASSOPTIONcompsoc
  \section*{Acknowledgments}
\else
  \section*{Acknowledgment}
\fi
Thank you to the anonymous Trollslayer crowd workers.

\bibliographystyle{IEEEtran}
\small{
\bibliography{paper-trolls}}
\end{document}